\documentclass[english,utf8]{frontiersSCNS}
\PassOptionsToPackage{vlined,ruled}{algorithm2e}
\usepackage[T1]{fontenc}
\usepackage[utf8]{inputenc}
\usepackage{color}
\usepackage{babel}
\usepackage{prettyref}
\usepackage{url}
\usepackage{algorithm2e}
\usepackage{amstext}
\usepackage{graphicx}
\usepackage[authoryear]{natbib}
\usepackage[unicode=true]
 {hyperref}

\makeatletter
\usepackage[T1]{fontenc}

\usepackage{xcolor}
\definecolor{darkblue}{rgb}{0.0,0.0,0.64}

\definecolor{choose_color}{HTML}{B0CFEC}
\definecolor{connect_color}{HTML}{EDAF9B}
\definecolor{rest_color}{HTML}{EBD7A0}

\usepackage{amsmath}

\usepackage[group-separator={,}, group-minimum-digits=4]{siunitx}

\usepackage{prettyref}
\newrefformat{chap}{\hyperref[#1]{Chapter~\ref*{#1}}}
\newrefformat{sec}{\hyperref[#1]{Section~\ref*{#1}}}
\newrefformat{subsec}{\hyperref[#1]{Section~\ref*{#1}}}
\newrefformat{apdx}{\hyperref[#1]{Appendix~\ref*{#1}}}
\newrefformat{lst}{\hyperref[#1]{Listing~\ref*{#1}}}
\newrefformat{fig}{\hyperref[#1]{Figure~\ref*{#1}}}
\newrefformat{tab}{\hyperref[#1]{Table~\ref*{#1}}}
\newrefformat{eqn}{\hyperref[#1]{Equation~\ref*{#1}}}

\usepackage{url,hyperref,lineno,microtype,subcaption}
\usepackage[onehalfspacing]{setspace}

\hypersetup{colorlinks=true, breaklinks=true, linkcolor=darkblue, urlcolor=darkblue, citecolor=darkblue}

\def\keyFont{\fontsize{8}{11}\helveticabold }
\def\firstAuthorLast{Pronold {et~al.}}

\def\Authors{Jari Pronold$^{1,2,*}$, Jakob Jordan\,$^{4}$, Brian J. N. Wylie\,$^{5}$, Itaru Kitayama\,$^{6}$, Markus Diesmann\,$^{1,3,7}$ and Susanne Kunkel\,$^{8}$}

\def\Address{
$^1$ Institute of Neuroscience and Medicine (INM-6) and Institute for Advanced Simulation (IAS-6) and JARA Institute Brain Structure-Function Relationships (INM-10), J\"ulich Research Centre, J\"ulich, Germany\\
$^2$ RWTH Aachen University, Aachen, Germany \\
$^3$ Department of Physics, Faculty 1, RWTH Aachen University, Aachen, Germany\\
$^4$ Department of Physiology, University of Bern, Bern, Switzerland\\
$^5$ J\"ulich Supercomputing Centre, J\"ulich Research Centre, J\"ulich, Germany\\
$^6$ RIKEN Center for Computational Science, Kobe, Japan\\
$^7$ Department of Psychiatry, Psychotherapy and Psychosomatics, Medical Faculty, RWTH Aachen University, Aachen, Germany\\
$^8$ Faculty of Science and Technology, Norwegian University of Life Sciences, {\AA}s, Norway
}

\def\corrAuthor{Jari Pronold}

\def\corrEmail{j.pronold@fz-juelich.de}

\makeatother

\begin{document}
\onecolumn

\global\long\def\B{\:\mathrm{B}}%
\global\long\def\kB{\:\mathrm{kB}}%
\global\long\def\MB{\:\mathrm{MB}}%
\global\long\def\GB{\:\mathrm{GB}}%
\global\long\def\TB{\:\mathrm{TB}}%
\global\long\def\PB{\:\mathrm{PB}}%
\global\long\def\MiB{\:\mathrm{MiB}}%
\global\long\def\GiB{\:\mathrm{GiB}}%
\global\long\def\TiB{\:\mathrm{TiB}}%
\global\long\def\PiB{\:\mathrm{PiB}}%
\global\long\def\GBps{\:\mathrm{GBps}}%
\global\long\def\bit{\:\mathrm{bit}}%
\global\long\def\bits{\:\mathrm{bits}}%
\global\long\def\GHz{\:\mathrm{GHz}}%
\global\long\def\Hz{\:\mathrm{Hz}}%
\global\long\def\rate{\nu}%
\global\long\def\indegree{K}%
\global\long\def\vps{MT}%
\global\long\def\numax{\nu_{\text{max}}}%
\global\long\def\NE{N_{\text{E}}}%
\global\long\def\NI{N_{\text{I}}}%
\global\long\def\E{\text{E}}%
\global\long\def\I{\text{I}}%
\global\long\def\pF{\mathrm{\:pF}}%
\global\long\def\mV{\mathrm{\:mV}}%
\global\long\def\pA{\mathrm{\:pA}}%
\global\long\def\s{\:\mathrm{s}}%
\global\long\def\ms{\:\mathrm{ms}}%
\global\long\def\us{\:\mu\text{sec}}%
\global\long\def\FLOPS{\:\mathrm{FLOPS}}%
\global\long\def\PFLOPS{\:\mathrm{PFLOPS}}%
\global\long\def\TFLOPS{\:\mathrm{TFLOPS}}%
\global\long\def\percent{\:\%}%
\global\long\def\NSEA{N_{\text{SEA}}}%
\global\long\def\NSWP{N_{\text{SWP}}}%

\firstpage{1}
\author[\firstAuthorLast ]{\Authors}  %This field will be automatically populated 
\address{}                            %This field will be automatically populated 
\correspondance{}                     %This field will be automatically populated
\extraAuth{}                          %If there are more than 1 corresponding author, comment this line and uncomment the next one. 
%\extraAuth{corresponding Author2 \\ Laboratory X2, Institute X2, Department X2, Organization X2, Street X2, City X2 , State XX2 (only USA, Canada and Australia), Zip Code2, X2 Country X2, email2@uni2.edu}
\title[Routing spikes by parallel sorting]{Routing brain traffic through the von Neumann bottleneck: Parallel sorting and refactoring}%
\phantomsection\addcontentsline{toc}{section}{Title}
\maketitle
\begin{abstract}

Generic simulation code for spiking neuronal networks spends the major
part of time in the phase where spikes have arrived at a compute node
and need to be delivered to their target neurons. These spikes were
emitted over the last interval between communication steps by source
neurons distributed across many compute nodes and are inherently irregular
and unsorted with respect to their targets. For finding those targets,
the spikes need to be dispatched to a three-dimensional data structure
with decisions on target thread and synapse type to be made on the
way. With growing network size a compute node receives spikes from
an increasing number of different source neurons until in the limit
each synapse on the compute node has a unique source. Here we show
analytically how this sparsity emerges over the practically relevant
range of network sizes from a hundred thousand to a billion neurons.
By profiling a production code we investigate opportunities for algorithmic
changes to avoid indirections and branching. Every thread hosts an
equal share of the neurons on a compute node. In the original algorithm
all threads search through all spikes to pick out the relevant ones.
With increasing network size the fraction of hits remains invariant
but the absolute number of rejections grows. \textcolor{black}{Our
new} alternative algorithm equally divides the spikes among the threads
and immediately sorts them in parallel according to target thread
and synapse type. After this every thread completes delivery solely
of the section of spikes for its own neurons. \textcolor{black}{Independent
of the number of threads, all spikes are looked at only twice.} The
new algorithm halves the number of instructions in spike delivery
which leads to a reduction of simulation time of up to $40\percent$.
Thus, spike delivery is a fully parallelizable process with a single
synchronization point and thereby well suited for many-core systems.
Our analysis indicates that further progress requires a reduction
of the latency \textcolor{black}{the} instructions experience in
accessing memory. The study provides the foundation for the exploration
of methods of latency hiding like software pipelining and software-induced
prefetching.

\keyFont{\section{Keywords:}spiking neural networks, large-scale
simulation, distributed computing, parallel computing, sparsity, irregular
access pattern, memory-access bottleneck}
\end{abstract}

\section{Introduction}

\label{sec:intro}

Over the last two decades simulation algorithms for spiking neuronal
networks have continuously been improved. The largest supercomputers
available can be employed to simulate networks with billions of neurons
at their natural density of connections. The respective codes scale
well over orders of magnitude of network size and number of compute
nodes \citep{Jordan18_2}. Still simulations \textcolor{black}{at
the brain scale} are an order of magnitude slower than real time,
hindering the investigation of processes like plasticity and learning
unfolding over hours and days of biological time. In addition, there
is a trend of aggregating more compute power in many-core compute
nodes. This further reduces the strain on inter-node communication
as one limiting component but increases the urgency to better understand
the fundamental operations required for routing spikes within a compute
node.

The spiking activity in mammalian neuronal networks is irregular,
asynchronous, sparse, and delayed. Irregular refers to the structure
of the spike train of an individual neuron. The intervals between
spike times are of different length and unordered as if drawn from
a random process. Consequently also the number of spikes in a certain
time interval appears random. Asynchronous means that the spikes of
any two neurons occur at different times and exhibit low correlation.
The activity of neurons is sparse in time as compared to the time
constants of neuronal dynamics; only few spikes are emitted in any
second of biological time. Lastly, there is a biophysical delay in
the interaction between neurons imposed by their anatomy. The delay
may be a fraction of a millisecond for neurons within a distance of
a few micrometers but span several milliseconds for connections between
brain regions \citep[see][ for an example compilation of parameters]{Schmidt18_1409}.

 The existence of a minimal delay in a network model together with
the sparsity of spikes has suggested a three-phase cycle for an algorithm
directly integrating the differential equations of the interacting
model neurons \citep{Morrison05a}. First, communication between compute
nodes occurs synchronously in intervals of the minimal delay. This
communication transmits all the spikes that have occurred on a compute
node since the last communication step to the compute nodes harboring
target neurons of these spikes. Second, the received spikes are delivered
to their target neurons and placed in spike ring buffers representing
any remaining individual delay. Finally, the dynamical state of each
neuron is propagated for the time span of the minimal delay while
the ring buffer is rotating accordingly. Once all neurons are updated
the next communication is due and the cycle begins anew.

Progress in each update phase is shaped by the spiking interaction
between neurons and independent of the level of detail of the individual
model neurons constituting the network. The choice of neuron model,
however, influences the distribution of computational load across
the phases of the simulation. Some studies require neuron models with
thousands of electrical compartments \citep{Markram2015_456} and
efficient simulation codes are available for this purpose \citep{Carnevale06,kumbhar2019coreneuron,Akar19_274}.
Here we focus on simulation code for networks of model neurons described
by a handful of differential equations as widely used in the computational
neuroscience community. These investigations range from studies with
several thousands of neurons on the fundamental interplay between
excitation and inhibition \citep{Brunel00_183} to models attempting
to capture the natural density of wiring \citep{Potjans14_785,Billeh20}
and the interaction between multiple cortical areas \citep{Schmidt18_e1006359,Joglekar18_222}.
Previous measurements on a production code \citep{Jordan18_2} already
show that for networks of such simple model neurons the dominating
bottleneck for further speed-up is neither the communication between
computes nodes nor the update of the dynamical state of the neurons,
but the spike-delivery phase. The empirical finding is elegantly confirmed
by an analytical performance model encompassing different types of
network and neuron models \citep{cremonesi2020analytic,cremonesi2020understanding}.
These authors further identify the latency of memory access as the
ultimate constraint of the spike-delivery phase.

Profiling tools like Intel VTune provide measures on where an application
spends its time and how processor and memory are used. Two basic measures
are the total number of instructions carried out and the number of
clockticks the processor required per instruction (CPI). The former
characterizes the amount of computations that need to be done to arrive
at the solution. The latter describes how difficult it is on average
to carry out an individual instruction due to the complexity of the
operation and the waiting for accessing the corresponding part of
memory. The product of the two measures is the total number of clockticks
and should correlate to the wall clock time required to complete the
simulation phase under investigation. Methods of software pipelining
and software-induced prefetching attempt to improve the CPI by better
vectorization of the code or by indicating to the processor which
memory block will soon be required. These optimizations may lead to
an increase in the actual number of instructions but as long as the
product with the reduced CPI decreases, performance is improving.
Nevertheless, a low CPI does not mean that the code is close to optimal
performance. If the code is overly complicated, for example by recalculating
known results or missing out on regularities in the data it may underutilize
data that has been retrieved from memory rendering advanced methods
of optimization fruitless. Therefore, in the present study, as a first
step, we do not consider pipelining and prefetching but exclusively
assess the number of instructions required by the algorithm. It turns
out that a better organized algorithm indeed avoids unnecessary tests
and indirections. This decrease in the number of instructions also
decreases CPI as a side effect until with increasing sparseness of
the network CPI climbs up again. The control flow in the code becomes
more predictable for the processor until the fragmentation of memory
limits the success. The results of our study give us some confidence
that further work can now directly address improving the CPI.

In \prettyref{sec:spike-delivery-as-mem-bottleneck} we expose spike
delivery as the present bottleneck for the simulation of \textcolor{black}{mammalian}
spiking neuronal networks, characterize analytically the transition
to sparsity with growing network size, and present the original algorithm
as well as state-of-the-art performance data. Next we introduce the
software environment of our study and the neuronal network model used
to obtain quantitative data (\prettyref{sec:benchmarking-framework}).
On the basis of these preparatory sections \prettyref{sec:redesign}
presents a new algorithm streamlining the routing of spikes to their
targets. Subsequently \prettyref{sec:results} evaluates the success
of the redesign and identifies the origin of the improvement by profiling.
Finally \prettyref{sec:discussion} embeds the findings into the ongoing
efforts to develop generic technology for the simulation of spiking
neuronal networks.

The conceptual and algorithmic work described here is a module in
our long-term collaborative project to provide the technology for
neural systems simulations \citep{Gewaltig_07_11204}. Preliminary
results have been presented in abstract form \citep{Kunkel19_ISC}.

\section{Spike delivery as memory-access bottleneck}

\label{sec:spike-delivery-as-mem-bottleneck}The temporally sparse
event-based communication between neurons presents a challenging memory-access
bottleneck in simulations of spiking neuronal networks for modern
architectures optimized for dense data. In our reference implementation
NEST (\prettyref{subsec:NEST}), delivery of spikes to their synaptic
and neuronal targets involves frequent access to essentially random
memory locations, rendering automatic prediction difficult and leading
to long data-access times due to ineffective use of caches. The following
subsection provides an analysis of the sparsity of the network representation
for increasing numbers of MPI processes and threads. Based on this,
there follows a description of the connection data structures and
spike-delivery algorithm in the original implementation. The final
subsection provides example benchmarking data for this state-of-the-art
simulation code.

\subsection{Sparsity of network representation}

\label{subsec:unique-sources}
\begin{figure}[!t]
\begin{centering}
\includegraphics{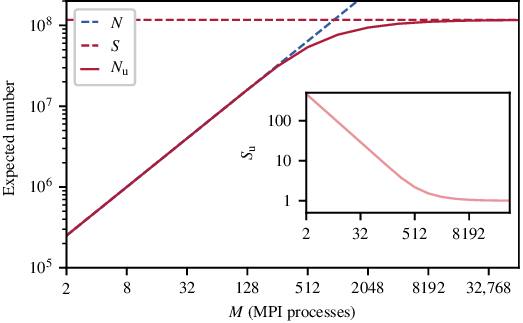}
\par\end{centering}
\caption{Expected number of unique source neurons $N_{\mathrm{u}}$ (pink curve)
of all thread-local synapses as a function of number of MPI processes
$M$ assuming $T=12$ threads and $125,000$ neurons per MPI process\textcolor{black}{{}
in a weak-scaling scenario}; total number of neurons $N$ (dashed
blue curve) and number of thread-local synapses $S$ (dashed pink
horizontal line for $K=11,250$ synapses per neuron). Inset: Expected
number of thread-local synapses per unique source neuron $S_{\mathrm{u}}=S/N_{\mathrm{u}}$
(light pink curve). All graphs in double logarithmic representation.\label{fig:unique-sources}}
\end{figure}
We consider a network of $N$ neurons distributed in a round-robin
fashion across $M$ MPI processes and $T$ threads per process. Each
neuron receives $K$ incoming synapses, which are represented on the
same thread as their target neuron. In a weak scaling scenario, the
computational load per process is kept constant. This implies that
the number of thread-local synapses 
\begin{equation}
S=NK/(MT)\label{eq:thread-local-synapses}
\end{equation}
 does not change. The total network size, in contrast, increases with
$MT$. In the limit of large network sizes each synapse on a given
thread originates from a different source neuron. This scenario was
already considered in \citep[section 2.4]{Kunkel14_78}, at the time
to analyze the increase in memory overhead observed with increasing
sparsity. For completeness here we briefly restate this result in
the parameters used in the present work.

The probability that a synapse has a particular neuron $j$ as source
neuron is $1/N$ and, conversely, the probability that the synapse
has a different source neuron is $1-1/N$. The probability that none
of the $S$ thread-local synapses has neuron $j$ as source is $p_{\mathrm{\emptyset}}=\left(1-1/N\right)^{S}$.
Conversely, $p=1-p_{\mathrm{\emptyset}}$ denotes the probability
that $j$ is source to at least one of the thread-local synapses.
Therefore, the expected number of unique source neurons of the thread-local
synapses is given by $N_{\mathrm{u}}=pN$ expanding to
\begin{equation}
N_{\mathrm{u}}=\left(1-\left[\left(1-\frac{1}{N}\right)^{N}\right]^{\frac{K}{MT}}\right)N
\end{equation}
which is equation (6) of \citep{Kunkel14_78}. In weak scaling $MT$
grows proportionally to $N$ such that
\[
N_{\mathrm{u}}=\left(1-\left[\left(1-\frac{1}{N}\right)^{N}\right]^{\frac{S}{N}}\right)N
\]
where they further identified the term $\left[\cdot\right]$ in the
limit of large $N$ as the definition of the exponential function
with argument $-1$ and therefore
\[
\widetilde{N}_{\mathrm{u}}=\left(1-\exp\left(-\frac{S}{N}\right)\right)N.
\]
They confirm that the limit of $N_{\mathrm{u}}$ is indeed $S$ and
that a fraction $\zeta$ of $S$ is reached at a network size of 
\begin{equation}
N_{\zeta}=\frac{S}{2\left(1-\zeta\right)}.
\end{equation}

\prettyref{fig:unique-sources} illustrates the point where in weak
scaling \textcolor{black}{the }total network size $N$ equals the
number of thread local synapses $S$. Here the number of unique source
neurons $N_{\mathrm{u}}$ of the thread-local synapses bends. According
to the definition \prettyref{eq:thread-local-synapses} of $S$, here
a particular target neuron chooses its $K$ incoming synapses from
the same total number of threads $MT=K$ and already half ($\zeta=\frac{1}{2}$)
of the source neurons of the thread-local synapses are unique. The
number of thread-local synapses per unique source neuron $S_{\mathrm{u}}$
indicates the sparsity of the network representation on a compute
node (inset of \prettyref{fig:unique-sources}). The measure converges
to one exhibiting a bend at the same characteristic point as $N_{\mathrm{u}}$.

\subsection{Memory layout of synapse and neuron representations}

\label{subsec:fundamental-data-structures}
\begin{figure}
\begin{centering}
\includegraphics{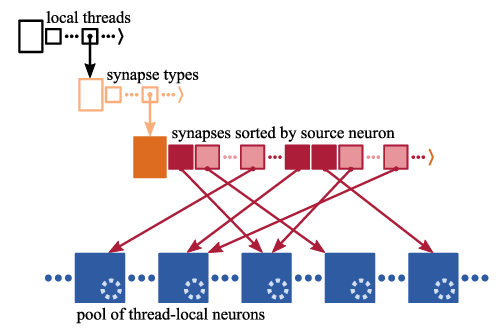}
\par\end{centering}
\caption{Memory layout of synapse and neuron representations on each MPI
process. Each process stores the local synapses (pink filled squares)
in a three-dimensional resizable array sorted by hosting thread and
synapse type. At the innermost level, synapses are arranged in source-specific
target segments (dark pink: first synapse; light pink: subsequent
targets); only one innermost array shown for simplicity. Target neurons
(blue filled squares) are stored in neuron-type and thread-specific
memory pools; only one pool shown for simplicity. Each neuron maintains
a spike ring buffer (dotted light blue circles). Synapses have access
to their target neurons through target identifiers (dark pink arrows).\label{fig:5g-data-structures}}
\end{figure}
A three-dimensional resizable array stores the process-local synapses
sorted by hosting thread and synapse type (\prettyref{fig:5g-data-structures}),
where synapses are small in size, each typically taking up few tens
of Bytes. Each synapse has access to its target neuron, which is hosted
by the same MPI process and thread \citep{Morrison05a}. The target
identifier provides access either through a pointer to the target
neuron consuming $8\B$ or an index of $2\text{\ensuremath{\B}}$
that is used to retrieve the corresponding pointer. Here we use the
latter implementation of the target identifier reducing per-synapse
memory usage at the cost of an additional indirection (see Section
3.3.2 in \citealp{Kunkel14_78}).

In the innermost arrays of the data structure, synapses are sorted
by source neuron, which is an optimization for small to medium scale
systems (see Section 3.3 in \citealp{Jordan18_2}) exploiting that
each neuron typically connects to many target neurons (out-degree).
Thereby, synapses are arranged in target segments, each consisting
of at least one target synapse potentially followed by subsequent
targets (\prettyref{subsec:unique-sources}). In a weak-scaling experiment
the increasing sparsity of the network in the small to medium scale
regime (\prettyref{subsec:unique-sources}) influences the composition
of the innermost array. As synapses are to an increasing degree distributed
across MPI processes and threads, the expected number of source-specific
target segments increases while the average segment size decreases
(cf. $N_{\text{\ensuremath{\mathrm{u}}}}$ and $S_{\mathrm{u}}$ in
\prettyref{fig:unique-sources}, respectively). Note that the degree
of distribution also depends on the number of synapse types, which
is however not considered in this study.

A model neuron easily takes up more than a Kilobyte of memory. Multi-chunk
memory pools enable contiguous storage of neurons of the same type
hosted by the same thread, where due to the many-to-one relation
between target synapses and neurons, the order of memory locations
of target neurons is independent of the order of synapses in the target
segments.

Synaptic transmission of spikes entails delays, which influence the
time when spikes take effect on the dynamics of the target neurons.
As typically synapses from many different source neurons converge
on the same target neuron (in-degree), it is more efficient to jointly
account for their delays in the neuronal target. Therefore, each neuron
maintains a spike ring buffer serving as temporary storage and scheduler
for the incoming spikes \citep{Morrison05a}.

\subsection{Original spike-delivery algorithm}

\label{subsec:original-deliver-algo}Every time all local neurons
have been updated and all recent spikes have been communicated across
MPI processes, the spike data needs to be delivered from the process-local
MPI receive buffers to the process-local synaptic and neuronal targets.
Each spike entry is destined for an entire target segment of synapses
(\prettyref{subsec:fundamental-data-structures}), which is an optimization
for the small to medium scale regime introduced in \citet{Jordan18_2}.
Therefore, each entry conveys the location of the target segment within
the three-dimensional data structure storing the process-local synapses
(\prettyref{fig:5g-data-structures}), i.e. identifiers for the hosting
thread and the type of the first synapse of the target segment, as
well as the synapse's index within the innermost resizable array.

In the original algorithm each thread reads through all spike entries
in the MPI receive buffer but it only proceeds with the delivery of
a spike if it actually hosts the spike's targets — all spike entries
indicating other hosting threads are skipped. Each thread delivers
the relevant spikes to every synapse of the corresponding target segments
one by one. On receiving a spike, a synapse transfers synaptic delay
and weight to the corresponding target neuron, where the stored target
identifier provides access to the neuron. The transmitted synaptic
properties, delay and weight, define time and amplitude of the spike's
impact on the neuron, respectively, allowing the neuron to add the
weight of the incoming spike to the correct position in the neuronal
spike ring buffer.

In a weak-scaling experiment the increasing sparsity of the network
in the small to medium scale regime (\prettyref{subsec:unique-sources})
influences algorithmic progression and memory-access patterns. Access
to target neurons and their spike ring buffers is always irregular
regardless of the degree of distribution of the network across MPI
processes, but memory access to synapses becomes progressively irregular.
The number of spike entries communicated via MPI increases to cater
to the growing number of target segments (\prettyref{subsec:fundamental-data-structures}).
In consequence, each thread needs to process ever more spike entries,
delivering relevant spikes to ever more but shorter target segments,
where successively visited target segments are typically in nonadjacent
memory locations. In the sparse limit where each target segment consists
of a single synapse, spike delivery to both neuronal and synaptic
targets requires accessing essentially random locations in memory.
As many synapses of different source neurons converge on the same
target neuron, it is impossible to arrange target neurons in memory
such that their order corresponds to the order in synaptic target
segments. The pseudocode in \prettyref{subsubsec:pseudocode-ORI}
summarizes this original spike-delivery algorithm.

\subsubsection{Pseudocode}

\label{subsubsec:pseudocode-ORI}
\begin{algorithm}
\SetAlgorithmName{}{}{}
\SetAlgoRefName{ORI nrn}
\DontPrintSemicolon
\SetKwProg{Fn}{}{}{}
\SetKwFunction{Receive}{Receive}
\SetKwFunction{AddValue}{AddValue}
\KwData{$spike\_ring\_buffer$}
\BlankLine
\Fn{\Receive{$delay$, $weight$}}{
  \nlset{RB}$spike\_ring\_buffer.$\AddValue{$delay$, $weight$}\;
}

\caption{\textcolor{black}{Original }\texttt{Receive()} procedure in neuron;
\textbf{\footnotesize{}RB} marks access to the spike ring buffer.\label{alg:ORI-nrn}}
\end{algorithm}
\begin{algorithm}
\SetAlgorithmName{}{}{}
\SetAlgoRefName{ORI syn}
\DontPrintSemicolon
\SetKwProg{Fn}{}{}{}
\SetKwFunction{Send}{Send}
\SetKwFunction{Receive}{Receive}
\SetKw{Return}{return}
\KwData{$subsq$, $target\_neuron$, $delay$, $weight$}
\BlankLine
\Fn{\Send{}}{
  $target\_neuron.$\Receive{$delay$, $weight$}\;
  \Return $subsq$\;
}

\caption{\textcolor{black}{Original }\texttt{Send()} function in synapse,
which calls the\texttt{ Receive()} procedure of the target neuron
(\ref{alg:ORI-nrn}) passing on synaptic properties.\label{alg:ORI-syn}}
\end{algorithm}
\begin{algorithm}
\SetAlgorithmName{}{}{}
\SetAlgoRefName{ORI}
\DontPrintSemicolon
\SetKwFunction{Send}{Send}
\SetKwFunction{GetTargetLoc}{GetTargetLoc}
\SetKw{In}{in}
\SetKw{True}{true}
\SetKw{TID}{\normalfont TID}
\KwData{$recv\_buffer$, $synapses$}
\BlankLine
\ForEach{$spike$ \In $recv\_buffer$}{
  $(tid,\,syn\_id,\,lcid) \leftarrow spike.$GetTargetLoc()\;
  \If{tid == \TID}{
    $subsq \leftarrow \True$\;
    \nlset{TS}\While{$subsq$}{
      \nlset{SYN}$subsq \leftarrow synapses[tid][syn\_id][lcid].$\Send{}\;
      $lcid \leftarrow lcid+1$\;
    }
  }
}

\caption{Original algorithm delivering spikes to local targets with support
for multi-threading, where $\mathrm{TID}$ denotes the identifier
of the executing thread. \textbf{\footnotesize{}TS} marks iteration
over a synaptic target segment.\textbf{ }\textbf{\footnotesize{}SYN}
marks access to an individual target synapse (\ref{alg:ORI-syn}).\label{alg:ORI}}
\end{algorithm}

The original algorithm delivers spikes to the neuronal spike ring
buffers through the target synapses. Each neuron owns a $spike\_ring\_buffer$,
where the neuron member function \texttt{Receive()} triggers the spike
delivery by calling the spike ring buffer member function \texttt{AddValue()},
which then adds the weight of the spike to the correct position in
the buffer (\prettyref{alg:ORI-nrn}; \textbf{\footnotesize{}RB}).
To this end, both \texttt{Receive()} and \texttt{AddValue()} require
the synaptic properties $delay$ and $weight$.

Each synapse stores properties such as $delay$ and $weight$, an
identifier enabling access to the target neuron ($target\_neuron$),
and an indicator ($subsq$) of whether the target segment continues
or not (\prettyref{alg:ORI-syn}). The synapse member function \texttt{Send()}
calls the member function \texttt{Receive()} of the target neuron
passing on the synaptic properties and returns the indicator $subsq$.

The original spike-delivery algorithm has access to the MPI spike-receive
buffer ($recv\_buffer$) containing all spike entries that need to
be delivered and to a three-dimensional resizable array of process-local
$synapses$ ordered by hosting thread and synapse type (\prettyref{alg:ORI};
see \prettyref{fig:5g-data-structures}). For each spike entry, the
3D location of the first target synapse is extracted and assigned
to the variables $tid$, $syn\_id$, and $lcid$, which indicate hosting
thread, synapse type, and location in the innermost $synapses$ array,
respectively. If the executing thread ($\mathrm{TID}$) is the hosting
thread of the target synapse, then the variable $lcid$ is used in
the enclosed while loop to iterate over the spike's entire synaptic
target segment within the innermost array $synapses[tid][syn\_id]$
(\textbf{\footnotesize{}TS}). To deliver a spike to the target synapse
at position $lcid$, the synapse member function \texttt{Send()} is
called on\break$synapses[tid][syn\_id][lcid]$ returning the indicator
$subsq$ (\textbf{\footnotesize{}SYN}).

\subsection{Simulation time}

\label{subsec:ORI-sim-time}
\begin{figure}
\begin{centering}
\includegraphics{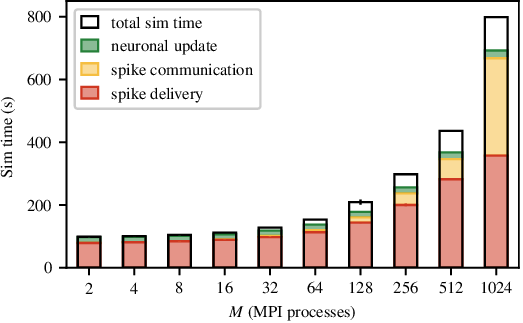}
\par\end{centering}
\caption{Contributions to simulation time (sim time) for spiking neural network
simulations with NEST (\prettyref{subsec:NEST}) where the number
of MPI processes increases proportionally with the total number of
neurons. Weak-scaling experiment running $2$ MPI processes per compute
node and $12$ threads per MPI process, with a workload of $125,000$
neurons per MPI process (network model see \prettyref{subsec:network-model}).
The network dynamics is simulated for $1\protect\s$ of biological
time; spikes are communicated across MPI processes every $1.5\protect\ms$.
Time spent on spike delivery (red bars), communication of spike data
(yellow bars), neuronal update (green bars), and total sim time (black
outline). Error bars (for most numbers of MPI processes hardly visible)
indicate the standard deviation over three repetitions. Timings obtained
via manual instrumentation of the respective parts of the source code;
measured on JURECA CM (\prettyref{subsec:systems}).\label{fig:intro-simtime-contributions}}
\end{figure}

The work of \citet{Jordan18_2} shows that spike delivery is the
dominating phase of simulation time from networks with a few hundred
thousand neurons to the regime of billions of neurons. In the latter,
the number of neurons in the network exceeds the number of synapses
represented on an individual compute node; each synapse on a given
compute node has a unique source neuron (\prettyref{subsec:unique-sources}).
Therefore a neuron finds either a single target neuron on a compute
node or none at all. Assuming a random distribution of neurons across
MPI processes, the network is fully distributed in terms of its connectivity.
From this point on the computational costs of spike delivery on a
compute node do not change with growing network size in a weak scaling
scenario; each synapse receives spikes with a certain frequency and
all spikes come from different sources. What is still increasing are
the costs of communication between the compute nodes. Nevertheless
for smaller networks below the limit of sparsity \citet{Jordan18_2}
provide optimizations (their section 3.3) exploiting the fact that
a spike finds multiple targets on a compute node. This reduces both
communication time and spike-delivery time, but the effect vanishes
in the limit (see Figure 7C in \citealp{Jordan18_2}; 5g-sort) where
the code continues to scale well with the maximal but invariant costs
of spike delivery.

The network model of \citet{Jordan18_2} exhibits spike-timing dependent
plasticity in its synapses between excitatory neurons. The spike-delivery
phase calculates the plastic changes at synapses because synaptic
weights only need to be known when a presynaptic spike is delivered
to its target \citep{Morrison07_1437}. Depending on the specific
plasticity rule these computations may constitute a considerable fraction
of the total spike delivery time. Therefore from the data of \citet{Jordan18_2}
we cannot learn which part of the spike-delivery time is due to the
calculation of synaptic plasticity and which part due to the actual
routing of spikes to their targets. In order to disentangle these
contributions the present study uses the same network model but considers
all synapses as static (\prettyref{subsec:network-model}).

\prettyref{fig:intro-simtime-contributions} shows a weak scaling
of our static neuronal network model across the critical region where
sparsity has not yet reached the limit. This confirms that even in
the absence of synaptic plasticity spike delivery is the dominant
contribution to simulation time independent of the number of MPI processes.
\textcolor{black}{The network on a single MPI process roughly corresponds
to the smallest cortical network in which the natural number of synapses
per neuron and the local connection probability of $0.1$ can simultaneously
be realized \citep{Potjans14_785}. While our weak scaling conserves
the former quantity, the latter drops.} In the regime from $2$ to
$512$ MPI processes the absolute time required for spike delivery
almost quadruples (factor of $3.9$). Beyond this regime, the relative
contribution of spike delivery to simulation time drops below $50\%$
because the time required for communication is increasing. The absolute
time for neuronal update remains unchanged throughout as the number
of neurons per MPI process is fixed.

\section{Benchmarking framework}

\label{sec:benchmarking-framework}

\subsection{Simulation engine}

\label{subsec:NEST}Over the past two decades simulation tools in
computational neuroscience have increasingly embraced a conceptual
separation of generic simulation engines and specific models of neuronal
networks \citep{Einevoll19_735}. Many different models can thus be
simulated with the same simulation engine. This enables the community
to separate the life cycle of a simulation engine from those of specific
individual models and to maintain and further develop simulation engines
as an infrastructure. Furthermore this separation is useful for the
cross-validation of different simulation engines.

One such engine is the open-source community code NEST\footnote{\url{https://www.nest-simulator.org}}
\citep{Gewaltig_07_11204}. The quantitative analysis of the state-of-the-art
in the present study is based on this code and alternative concepts
are evaluated in its software framework. This ensures that ideas are
immediately exposed to the complications and legacy of real-world
code. NEST uses a hybrid between an event-driven and a time-driven
simulation scheme to exploit that individual synaptic events are rare
whereas the total number of spikes arriving at a neuron is large \citep{Morrison05a}.
Neurons are typically updated every $0.1\ms$ and spike times are
constrained to this time grid. For high-precision simulations spikes
can also be processed in continuous time \citep{Morrison07_47,Hanuschkin10_113}.
In contrast, synapses are only updated when a spike is arriving from
the corresponding presynaptic neuron. The existence of a biophysical
delay in the spiking interaction between neurons enables a global
exchange of spike data between compute nodes in intervals of the minimal
delay. The data structures and algorithms for solving the equations
of neuronal networks of natural size are documented and discussed
in the literature in detail \citep{Morrison08_267,Helias12_26,Kunkel2012_5_35,Kunkel14_78,Jordan18_2}
as well as technology for network creation \citep{Ippen2017_30} and
the language interface \citep{Eppler09_12,Zaytsev14_23,Plotnikov16_93}.
For the purpose of the present study it suffices to characterize the
main loop of state propagation (\prettyref{sec:intro}) and concentrate
on the details of spike delivery (\prettyref{sec:spike-delivery-as-mem-bottleneck}).

Besides spikes NEST supports gap junctions \citep{Hahne15_00022,Jordan20_12}
as a further biophysical mechanism of neuronal interaction. To allow
modeling of mechanisms affecting network structure on longer time
scales, NEST implements models of neuromodulated synaptic plasticity
\citep{Potjans10_103389}, voltage-dependent plasticity \citep{Stapmanns21_609147},
and structural plasticity \citep{Diaz16_57}. For the representation
of more abstract network models, NEST in addition supports binary
neuron models \citep{Grytskyy13_131} and continuous neuronal coupling
\citep{Hahne17_34} for rate-based and population models.

The present work is based on commit 059fe89 of the NEST 2.18 release.

\subsection{Network model}

\label{subsec:network-model}As earlier studies on neuronal network
simulation technology \citep[latest][]{Jordan18_2}, we use a \textcolor{black}{generic
model of mammalian neuronal} network\textcolor{black}{s} \citep{Brunel00_183}
for measuring and comparing proposed algorithmic modifications. The
model description and parameters are given in parameter tables \textcolor{black}{1–3
}\texttt{}of \citet{Jordan18_2} and \prettyref{subsec:ORI-sim-time}
gives an overview of performance for state-of-the-art code. A figure
\textcolor{black}{illustrating the structure of the model is part
of the NEST user-level documentation}\footnote{\textcolor{black}{\url{https://nest-simulator.readthedocs.io}}}.
The sole difference of the investigated model with respect to previous
studies is the restriction to static synapses for excitatory-excitatory
connections. These synapses have a fixed weight whereas in former
studies they exhibited spike-timing dependent plasticity \citep{Morrison07_1437}.

The network is split into two populations: excitatory ($80\%$) and
inhibitory neurons ($20\%$). These are modeled by single-compartment
leaky-integrate-and-fire dynamics with alpha-shaped postsynaptic currents.
Parameters are homogeneous across all neurons. Each neuron receives
a fixed number of excitatory and inhibitory connections with presynaptic
partners randomly drawn from the respective population. Thus every
neuron has $11,250$ incoming and, on average, $11,250$ outgoing
synapses, independent of the network size. Inhibitory synapses are
stronger than the excitatory synapses to ensure stability of the dynamical
state of the network. The simulation of $10\ms$ of biological time,
called init phase, is followed by the further simulation of $1\s$
of biological time. The former initiates the creation and initialization
of data structures which are unchanged in the simulation of subsequent
time stretches. The measured wall-clock time of the latter, called
simulation phase, is referred to as ``sim time''. The mean firing
rate across all network sizes considered in this study is $7.56\Hz$
with a standard deviation of $0.1\Hz$.

\subsection{Systems}

\label{subsec:systems}The JURECA Cluster Module (JURECA CM) and
the K computer are already specified in \citet{Jordan18_2}, their
characteristics are repeated here in the same words for completeness
except the renaming of JURECA to JURECA CM after the addition of a
booster module not used here. JURECA CM \citep{Krause:850758} consists
of $1872$ compute nodes, each housing two Intel Xeon E5-2680 v3 Haswell
CPUs with 12 cores each at $2.5\GHz$ for a total of $1.8\PFLOPS$.
Most of the compute nodes have $128\GB$ of memory available. In
addition $75$ compute nodes are equipped with two NVIDIA K80 GPUs,
which, however, are not used in this study. The nodes are connected
via Mellanox EDR Infiniband.

DEEP-EST\footnote{\url{https://www.deep-projects.eu}} (Dynamical
Exascale Entry Platform - Extreme Scale Technologies) is an EU project
exploring the usage of modular supercomputing architectures. Among
other components it contains a cluster module (DEEP-EST CM) tuned
for applications requiring high single-thread performance and a modest
amount of memory. The module consists of one rack containing $50$
nodes, each node hosting two Intel Xeon Gold $6146$ Skylake CPUs
with $12$ cores each. The CPUs run at $3.2\GHz$ and have $192\GB$
RAM. In total, the system has $45\TFLOPS$ and aggregates $45\TB$
of main memory. The system uses Mellanox InfiniBand EDR ($100\GBps$)
with fat tree topology for communication.

Both on JURECA CM and DEEP-EST CM we compile the application with
OpenMP enabled using GCC and link against ParaStationMPI for MPI support.
In our benchmarks, to match the node architecture, we launch $2$
MPI processes each with $12$ threads on every node and bind the MPI
processes to sockets using \texttt{-{}-cpu\_bind=sockets} to ensure
that the threads of each process remain on the same socket.

The K computer \citep{Miyazaki12} features $82,944$ compute nodes,
each equipped with an $8$-core Fujitsu SPARC64 VIIIfx processor operating
at $2\GHz$, with $16\GB$ RAM per node, leading to a peak performance
of about $11.3\PFLOPS$ and a total of $1377\TB$ of main memory.
The compute nodes are interconnected via the ``Tofu'' (``torus
connected full connection'') network with $5\GBps$ per link. The
K computer supports hybrid parallelism with OpenMP (v3.0) at the single
node level and MPI (v2.1) for inter-node communication. Applications
are compiled with the Fujitsu C/C++ Compiler and linked with Fujitsu
MPI. Each node runs a single MPI process with 8 threads.

\subsection{Software for profiling and workflow management}

\label{subsec:JUBE-VTune}Optimizing software requires the developer
to identify critical sections of the code and to guarantee identical
initial conditions for each benchmark. This is all the more true in
the field of simulation technology for spiking neuronal networks.
Despite the advances \citep{Schenck14_SC,cremonesi2019computational,cremonesi2020analytic,cremonesi2020understanding}
in the categorization of neuronal network applications and the identification
of bottlenecks, performance models are not yet sufficiently quantitative
and fundamental algorithms and data structures are evolving. Therefore
the field still relies on exploration and quantitative experiments.
The present work employs the profiling tool VTune to guide the development
as well as the \textcolor{black}{benchmarking environment}\texttt{}
JUBE\textcolor{black}{{} for workflow management}\texttt{}. In addition
the NEST code contains manual instrumentation to gather the cumulative
times spent in the update, communicate, and deliver phases and to
determine the total simulation time.

\subsubsection{VTune}

\label{subsec:VTune}VTune Profiler\footnote{\url{https://software.intel.com/vtune}}
is a proprietary performance analysis tool developed by the company
Intel providing both a graphical user interface and a command line
interface. It collects performance statistics across threads and MPI
processes while the application is running. VTune supports different
analysis types instructing the profiling program executing the application
to focus on specific characteristics. From the rich set of statistical
measures we select only three basic quantities: Instructions Retired,
Clockticks, and Clockticks per Instructions Retired (CPI). The Instructions
Retired show the total number of completed instructions, while the
CPI is the ratio of unhalted processor cycles (clockticks) relative
to the number of instructions retired indicating the impact of latency
on the application's execution.

\subsubsection{JUBE}

Documenting and reproducing benchmarking data requires the specification
of metadata on the computer systems addressed and metadata on the
configurations for compiling the application, for running the simulations,
and for evaluating the results. The Jülich Benchmarking Environment
(JUBE) \footnote{\url{https://www.fz-juelich.de/jsc/jube}}\citep{Luehrs16_432}
is a software suite developed by the Jülich Supercomputing Centre.
We employ JUBE to represent all metadata of a particular benchmark
by a single script.

\section{Redesign of spike-delivery algorithm}

\label{sec:redesign}

\label{subsec:prep-deliver}The algorithmic redesign concentrates
on the initial part of spike delivery and access to the spike ring
buffers. The initial part of the original algorithm (\prettyref{subsec:original-deliver-algo})
does not fully parallelize the sorting of spike events according to
target thread (\prettyref{subsec:spike-receive-register}). Furthermore,
access to the spike ring buffers is hidden from the algorithm as the
buffer is considered an implementation detail of the object representing
a neuron \textcolor{black}{(}\prettyref{subsec:pointer-to-rb}\textcolor{black}{)}\texttt{}.
Acronyms given in the titles of the subsections label the specific
modifications for brevity and serve as references in pseudocode and
figures.

\subsection{Streamlined processing of spike entries (SRR)}

\label{subsec:spike-receive-register}In the original spike-delivery
algorithm (\prettyref{subsec:original-deliver-algo}), each thread
needs to read all spike entries in the MPI receive buffer, even those
not relevant for its thread local targets, causing an overhead per
spike entry, and hence per process-local synaptic target segment.
Moreover, for each relevant spike entry the thread hosting the targets
needs to identify the correct innermost array in the three-dimensional
resizable array storing the process-local synapses (\prettyref{fig:5g-data-structures})
based on the synapse-type information provided by the spike entry.
This entails additional per target-segment overhead.

To address these issues, we adapt the original spike-delivery algorithm
such that instead of directly dispatching the data from the receive
buffer to the thread-local targets, we introduce a two-step process:
First, the threads sort the spike entries by hosting thread and synapse
type in parallel, and only then the threads dispatch the spikes, now
exclusively reading relevant spike entries. To this end, we introduce
a new data structure of nested resizable arrays, called spike-receive
register (SRR), where each thread is assigned its own domain for writing.
After each spike communication, a multi-threaded transfer of all spike
entries from the MPI receive buffer to the spike-receive register
takes place: Each thread reads a different section of the entire receive
buffer and transfers the entries to their SRR domains. The domains
are in turn organized into separate resizable arrays, one per hosting
thread. Nested resizable arrays enable the further sorting by synapse
type. In this way each element of the MPI receive buffer is read only
once and spike entries are immediately sorted. This allows for a subsequent
multi-threaded delivery of spikes from the spike-receive register
to the corresponding target synapses and neurons such that all spike
entries are exclusively read by their hosting thread. At this point,
all a hosting thread has to do is to sequentially work through every
resizable array exclusively prepared for it in the sorting phase.
The additional sorting by synapse type allows the hosting thread to
deliver all spikes targeting synapses of the same type in one pass.

\subsection{Exposure of code dependencies (P2RB)}

\label{subsec:pointer-to-rb}In the original spike-delivery algorithm
(\prettyref{subsec:original-deliver-algo}), the target synapse triggers
the delivery of a spike to its target neuron, which then adds the
spike to its spike ring buffer. For the entire spike-delivery process,
this results in alternating access to target synapses and target neurons,
or more precisely, the target neurons' spike ring buffers. As synapses
store the target identifiers and other relevant information, access
to a target synapse is a precondition for access to its target neuron.

In order to expose this code dependency, we separate the two delivery
steps: spike delivery to target synapse and corresponding target neuron
are now triggered sequentially at the same call-stack level. Moreover,
instead of storing a target identifier, each synapse now stores a
pointer to the target neuron's spike ring buffer allowing for direct
access when delivering a spike. Therefore, the quantitative analysis
(\prettyref{subsec:results-preparatory-refactoring}) refers to this
set of optimizations as P2RB as an acronym for ``pointer to ring
buffer''.

\subsection{Pseudocode}

\label{subsec:pseudocode-SRR+P2RB}
\begin{algorithm}
\SetAlgorithmName{}{}{}
\SetAlgoRefName{SRR+P2RB syn}
\DontPrintSemicolon
\SetKwProg{Fn}{}{}{}
\SetKwFunction{Send}{Send}
\SetKw{Return}{return}
\KwData{$subsq$, $target\_rb$, $delay$, $weight$}
\BlankLine
\Fn{\Send{}}{
  \Return $(subsq,\,target\_rb,\,delay,\,weight)$\;
}

\caption{Adapted \texttt{Send()} function in synapse, which returns the pointer
to the spike ring buffer of the target neuron ($target\_rb$) owned
by the synapse and the synaptic properties required for spike delivery
to the target neuron.\label{alg:SRR+P2RB-syn}}
\end{algorithm}
\begin{algorithm}
\SetAlgorithmName{}{}{}
\SetAlgoRefName{SRR+P2RB}
\DontPrintSemicolon
\SetKwFor{ParForEach}{parallel foreach}{do}{}
\SetKwFunction{Send}{Send}
\SetKwFunction{GetTargetLoc}{GetTargetLoc}
\SetKwFunction{PushBack}{PushBack}
\SetKwFunction{AddValue}{AddValue}
\SetKw{In}{in}
\SetKw{To}{to}
\SetKw{True}{true}
\SetKw{TID}{\normalfont TID}
\SetKw{MAXTID}{\normalfont MAX\_TID}
\SetKw{MAXSYN}{\normalfont MAX\_SYN\_ID}
\KwData{$recv\_buffer$, $synapses$, $spike\_reg$}
\BlankLine
\ParForEach{$spike$ \In $recv\_buffer$}{
  $(tid,\,syn\_id,\,lcid) \leftarrow spike.$GetTargetLoc()\;
  $spike\_reg[\TID][tid][syn\_id].$\PushBack{$spike$}\;
}
\For{$syn\_id \leftarrow 0$ \To \MAXSYN}{
  \For{$tid \leftarrow 0$ \To \MAXTID}{
    \ForEach{$spike$ \In $spike\_reg[tid][\TID][syn\_id]$}{
      $lcid \leftarrow spike.lcid$\;
      $subsq \leftarrow \True$\;
      \nlset{TS}\While{$subsq$}{
        \nlset{SYN}$(subsq,\,target\_rb,\,d,\,w) \leftarrow synapses[\TID][syn\_id][lcid].$\Send{}\;
        $lcid \leftarrow lcid+1$\;
        \nlset{RB}$target\_rb.$\AddValue{$d$, $w$}\;
      }
    }
  }
}

\caption{Detailed reference algorithm delivering spikes to local targets with
support for multi-threading, where $\mathrm{TID}$ denotes the identifier
of the executing thread. \textbf{\footnotesize{}TS} marks iteration
over a synaptic target segment.\textbf{ }\textbf{\footnotesize{}SYN}
marks access to an individual target synapse (\ref{alg:SRR+P2RB-syn});
\textbf{\footnotesize{}RB} marks access to the spike ring buffer of
the corresponding target neuron. Based on \prettyref{alg:ORI}.\label{alg:SRR+P2RB-detailed}}
\end{algorithm}

The pseudocode \prettyref{alg:SRR+P2RB-detailed} illustrates the
changes to the original spike-delivery algorithm (\prettyref{alg:ORI})
resulting from the two new algorithms SRR (\prettyref{subsec:spike-receive-register})
and P2RB (\prettyref{subsec:pointer-to-rb}).

Instead of a target-neuron identifier, each synapse now owns a pointer
($\ensuremath{target\_rb}$) to the neuronal spike ring buffer. The
synapse member function \texttt{Send()} returns the pointer and the
synaptic properties $delay$ and $weight$ in addition to the indicator
$subsq$ (\prettyref{alg:SRR+P2RB-syn}). This allows the algorithm
to directly call \texttt{AddValue()} on the spike ring buffer (\prettyref{alg:SRR+P2RB-detailed};
\textbf{\footnotesize{}RB}) after the call to the synapse member function
\texttt{Send()} (\textbf{\footnotesize{}SYN}). The \texttt{Receive()}
member function of the target neuron (\prettyref{alg:ORI-nrn}) is
no longer required. Additionally, the algorithm now makes use of a
spike receive register ($\ensuremath{spike\_reg}$) for a preceding
thread-parallel sorting of the spike entries from the MPI receive
buffer by hosting thread ($tid$) and synapse type ($syn\_id$), where
each thread writes to its private region of the register ($\ensuremath{spike\_reg}[\mathrm{TID}]$).
Spikes are then delivered from the spike receive register instead
of the MPI receive buffer, where each thread processes only those
regions of the register that contain spike entries for thread-local
targets ($spike\_reg[tid][\mathrm{TID}]$ for all possible $tid$).

\section{Results}

\label{sec:results} The new data structures and algorithms of
\prettyref{sec:redesign} can be combined because they modify different
parts of the code. As the efficiency of the optimizations may depend
on the hardware architecture we assess their performance on three
computer systems (\prettyref{subsec:results-preparatory-refactoring}).
Subsequently we investigate \textcolor{black}{in }\texttt{}\prettyref{subsec:results_vtune}
the origin of the performance gain by evaluating the change in the
total number of instructions required and the average number of clockticks
consumed per instruction.

\subsection{Effect of redesign on simulation time}

\label{subsec:results-preparatory-refactoring}

\begin{figure}[!t]
\begin{centering}
\includegraphics{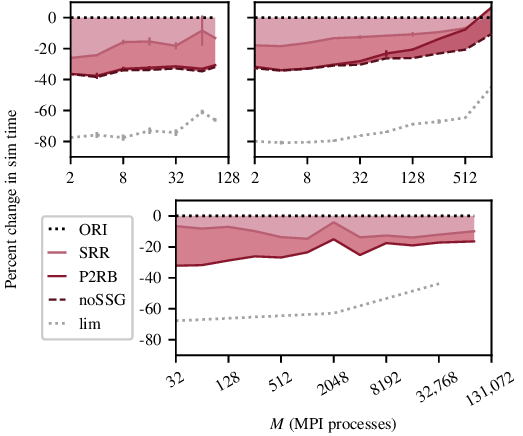}
\par\end{centering}
\caption{Cumulative relative change in simulation time after redesign of spike
delivery as a function of the number of MPI processes $M$. Top left
panel DEEP-EST CM and top right panel JURECA CM: linear-log representation
for number of MPI processes $M\in\{2;\:4;\:8;\:16;\:32;\:64;\:90\}$
and $M\in\{2;\:4;\:8;\:16;\:32;\:64;\:128;\:256;\:512;\:1024\}$,
respectively. Weak scaling of benchmark network model with same configuration
as in \prettyref{fig:intro-simtime-contributions}; error bars show
standard deviation based on $3$ repetitions. Bottom panel K computer:
number of MPI processes $M\in\{32;\:64;\:128;\:256;\:512;\:1024;\:2048;\:4096;\:8192;\:16,384;\:32,768;\:82,944\}$,
gray dotted curve: $M\in\{32;\:2048;\:32,768\}$. Weak scaling with
different configuration ($1$ MPI process per compute node; $8$ threads
per MPI process; $18,000$ neurons per MPI process). Black dotted
line at zero indicates performance of original code (ORI, \prettyref{subsec:original-deliver-algo}).
Light carmine red curve indicates change in sim time (shading fills
area to reference) due to sorting of spike entries prior to delivery
(SRR, \prettyref{subsec:spike-receive-register}). Dark carmine red
curve indicates additional change in sim time due to providing synapses
with direct pointers to neuronal spike ring buffers (P2RB, \prettyref{subsec:pointer-to-rb}).
Dashed brown curve shows additional change in sim time after removal
of an unrequired generic function call (noSSG). Gray dotted curve
indicates hypothetical limit to the decrease in sim time assuming
spike delivery takes no time.\label{fig:results-SRR-P2RB}}
\end{figure}
We select three computer systems for their differences in architecture
and size (\prettyref{subsec:systems}) to measure simulation times
for a weak scaling of the benchmark network model (\prettyref{subsec:network-model}).
The number of neurons per MPI process is significantly larger on the
DEEP-EST CM and the JURECA CM ($125,000$) than on the K computer
($18,000$) making use of the respectively available amount of memory
per process. On all three systems, we observe a relative reduction
in simulation time by more than $30\%$ (\prettyref{fig:results-SRR-P2RB})
for the combined optimizations compared to the original code (ORI,
\prettyref{subsec:original-deliver-algo}). This includes the removal
of a call\textcolor{black}{{} to a function named} \texttt{\textcolor{black}{set\_sender\_gid()}}\texttt{}
from the generic spike delivery code (noSSG). This function attaches
identifying information about the source of the corresponding spike
which is only required by specific non-neuronal targets such as recorders.
However, it causes per-target-segment overhead in all simulations.
The functionality can hence be moved to a more specialized part of
the code, e.g., the recorder model, and thereby regained if required.
The DEEP-EST CM hardly benefits from the removal of the call but the
batchwise processing of target segments has an increasing gain reaching
$20\%$ at $90$ MPI processes. On JURECA CM the function call does
limit the performance and its removal alone improves the performance
by $20\%$ for large numbers of MPI processes. Across systems and
number of MPI processes, the combined optimizations lead to a sustained
reduction in simulation time.

The new data structures and algorithms address the spike-delivery
phase only, but an optimization can only reduce simulation time to
the extent the component of the code to be optimized contributes to
the total time consumed as indicated by the limiting curve in \prettyref{fig:results-SRR-P2RB}.
In the neuronal network simulations considered here, the delivery
of spikes from MPI buffers to their targets consumes the major part
of simulation time. Initially, spike delivery takes up $80\%$ of
the simulation time for the DEEP-EST CM and the JURECA CM and $70\%$
for the K computer, but on all three systems the relative contribution
decreases with increasing number of MPI processes. Under weak scaling
into regimes beyond $1024$ MPI processes, the absolute time required
for spike delivery also initially grows but converges as the expected
number of thread-local targets per spike converges to one \citep[cf.][]{Jordan18_2}.
Although spike-delivery time increases throughout the entire range
of MPI processes on DEEP-EST CM and JURECA CM, the relative contribution
to simulation time declines because the time required by communication
between MPI processes increases more rapidly (for JURECA \textcolor{black}{CM
}\texttt{}data cf. \prettyref{fig:intro-simtime-contributions}).

P2RB increases the size of synapse objects by introducing an $8\B$
pointer to the neuronal spike ring buffer replacing the $2\B$ local
neuron index of the original algorithm (\prettyref{subsec:original-deliver-algo}).
We hypothesize that this increase underlies the declining success
and ultimately disadvantageous effect observed on JURECA CM. Control
simulations using the original code but with an artificially increased
object size confirm this hypothesis (data not shown).

\subsection{Origin of improvement}

\label{subsec:results_vtune}
\begin{figure}[!t]
\begin{centering}
\includegraphics{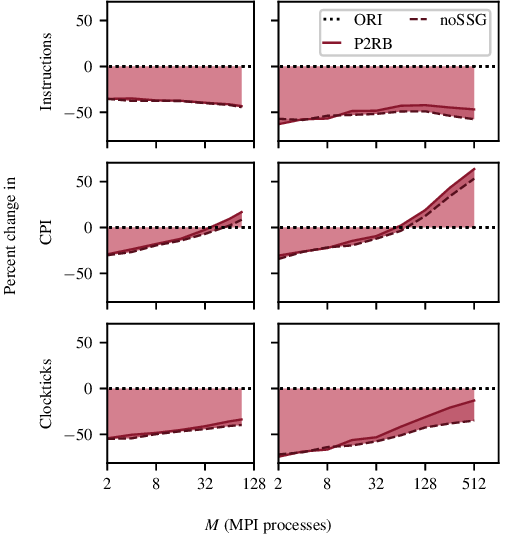}
\par\end{centering}
\caption{Relative change in instructions retired (top row), clockticks per
instruction retired (CPI, middle), and clockticks (bottom) during
spike delivery for P2RB (including SRR as in \prettyref{fig:results-SRR-P2RB})
and noSSG as a function of the number of MPI processes. Raw data for
all three quantities obtained by VTune (\prettyref{subsec:VTune}).
Left column DEEP-EST CM and right column JURECA CM: linear-log representation
for number of MPI processes $M\in\{2;\:4;\:8;\:16;\:32;\:64;\:90\}$
and $M\in\{2;\:4;\:8;\:16;\:32;\:64;\:128;\:256;\:512\}$, respectively.
Black dotted line at zero percent (\prettyref{alg:ORI}, \prettyref{subsec:original-deliver-algo})
indicates performance of the original code. Weak scaling of benchmark
network model as in \prettyref{fig:results-SRR-P2RB}.\label{fig:vtune}}
\end{figure}
The new data structures and algorithms realize a more fine-grained
parallelization and avoid indirections in memory accesses (\prettyref{sec:redesign}).
These changes significantly speed up the application (\prettyref{fig:results-SRR-P2RB})
across architectures and network sizes. In order to understand the
origin of this improvement we employ the profiling tool VTune (\prettyref{subsec:VTune})
which gives us access to the CPU's microarchitectural behavior. In
the analysis we concentrate on the total number of instructions executed
and the clockticks per instruction retired (CPI).

The total number of instructions decreases by close to $50\%$ on
all scales. Nevertheless, the contribution of noSSG to the reduction
in the number of instructions becomes larger as this algorithm removes
code which is called for every target segment. The number of target
segments, however, increase\textcolor{black}{s} with the number of
MPI processes until a limit is asymptotically approached (\prettyref{fig:unique-sources}).

For small problem sizes the CPI decreases when compared against the
baseline (\prettyref{alg:SRR+P2RB-detailed}), but at around $32$
MPI processes the instructions start to consume more clockticks than
in the original algorithm (\prettyref{fig:vtune}). This behavior
is apparent on DEEP-EST CM as well as JURECA CM where the additional
noSSG optimizations improves performance slightly. We interpret this
observation as follows. Initially the more orderly organization of
memory enables a shorter latency in memory access. At larger network
sizes CPI is dominated by memory access to the fragmented target segments
and this dominance is more pronounced as the new code spends fewer
instructions on reading the receive buffer.

Taken individually, the two metrics alone are not sufficient for explaining
the decrease in simulation time (\prettyref{fig:results-SRR-P2RB}).
The product of the number of instructions retired and CPI expresses
their interplay and reduces to the total number of clockticks required.
Thus, this product is a quantity that directly relates to the separately
measured sim time. Indeed, the comparison of this measure, depicted
in \prettyref{fig:vtune}, with \prettyref{fig:results-SRR-P2RB}
shows that the product qualitatively explains the change in sim time.
While CPI increases beyond baseline, the growth in sim time is slowed
down by having fewer instructions in total.

\section{Discussion}

\label{sec:discussion}Our investigation characterizes the dominance
of the spike-delivery phase in a weak-scaling scenario for a typical
random network model (\prettyref{subsec:ORI-sim-time}). At small
to medium network sizes spike delivery is the sole major contributor
to simulation time. Only if thousands of compute nodes are involved
communication between nodes becomes prominent (\prettyref{fig:intro-simtime-contributions})
while spike delivery remains the largest contributor \citep{Jordan18_2}.
The absolute time spent on spike delivery \textcolor{black}{is dominating
for small networks and} grows with increasing network size. The reason
for this is that the random network under study takes into account
that in the mammalian brain a neuron can send spikes to more than
ten-thousand targets. With increasing network size, these targets
are distributed over more and more compute nodes until in the limit
a neuron either finds a single target on a given compute node or more
likely none at all. As the number of synapses a compute node represents
is invariant under weak scaling, the node needs to process an increasing
number of spikes from different source neurons. For the simulation
parameters in this study the expected number of unique source neurons
and thereby absolute costs approach the limit when the network is
distributed across thousands of compute nodes (\prettyref{subsec:unique-sources})
thus also limiting the costs of spike delivery.

In spike delivery a thread inspects all spikes arriving at the compute
node. If the thread hosts at least one target neuron of a spike, the
thread needs to access a three-dimensional data structure (\prettyref{fig:5g-data-structures})
to activate the corresponding synapses and ultimately under consideration
of synapse specific delays place the spike in the ring buffers of
the target neurons. The present work investigates whether an alternative
algorithm can reduce the number of instructions and decisions when
handling an individual spike. The hypothesis is that a more compact
code and more predictable control flow allows modern processors a
faster execution. On purpose no attempt is made to apply techniques
like hardware prefetching or software pipelining to conceptually separate
improvements of the logic of the algorithm from further optimizations
that may have a stronger processor dependence. Nevertheless, we hope
that the insights of the present work provide the basis for any future
exploration of these issues.

The first step in our reconsideration of the spike-delivery algorithm
is to look at the initial identification of the relevant spikes for
each thread. Originally, each thread inspects all spikes. This means
that the algorithms perform many read operations on spikes without
further actions and that their proportion increases with an increasing
number of threads per compute node. The alternative algorithm which
we refer to as SRR (\prettyref{subsec:spike-receive-register}) carries
out a partial sorting of the spikes. Each thread is responsible for
an equally sized chunk of the incoming spikes and sorts them into
a data structure according to the thread on which the target neuron
resides and according to the type of the target synapse. Once all
threads have completed their work, they find a data structure containing
only relevant spikes and complete the spike delivery entirely independent
from the other threads. This already leads to a reduction of simulation
time between $10$ and $20\percent$ on the three computer systems
tested while the detailed development of this fraction differs with
network size.

As a second step we remove an indirection originating in the initial
object oriented design of the simulation code. Following the concept
of describing entities of nature by software objects, neurons became
objects receiving and emitting spikes and neuronal spike ring buffers
an implementation detail of no relevance for other components. As
a consequence when a neuron object receives a spike it needs to decide
in which ring buffer to place the spike, for example to separate excitatory
from inhibitory inputs, and delegate this task to the respective buffer.
Our alternative algorithm (P2RB, \prettyref{subsec:pointer-to-rb})
exposes the corresponding spike ring buffer to the synapse at the
time of network construction. The synapse stores the direct pointer
and no further decision is required during simulation. This change
further reduces simulation time by $10$ to $20\percent$.

One computer system (JURECA CM) shows a pronounced decline in the
computational advantage of the combined new algorithm (SRR+P2RB) for
large network sizes, which in case of P2RB we assume to be due to
an increase in synaptic memory footprint. An additional optimization
removing a generic function call that enriches spike events by information
on the identity of the source neuron mitigates the loss in performance.
As this functionality is not required for the interaction between
neurons we moved the function to a more specialized part of the code
(noSSG, \prettyref{subsec:results-preparatory-refactoring}).

The achievement of the combined algorithm (SRR+P2RB+noSSG) needs to
be judged in the light of the potential maximum gain. For small networks
spike delivery consumes $70$ to $80\percent$ of simulation time,
depending on the computer system, while this relative contribution
declines with growing network sizes as communication becomes more
prominent. Thus, the streamlined processing of spikes reduces spike
delivery by $50\percent$ largely independent of network size. In
conclusion, with the new algorithm spike delivery still substantially
contributes to simulation time.

In the small to medium scale regime (DEEP-EST CM, JURECA CM) the new
code gains its superiority from executing only half of the number
of instructions of the original implementation (\prettyref{subsec:results_vtune}).
The reduction becomes slightly larger with increasing network size.
This is plausible as for a given thread, the algorithm avoids processing
of a growing number of irrelevant spikes (SRR). As neurons have a
decreasing number of targets on a compute node while the total number
of synapses on the compute node stays invariant also the number of
relevant spikes increases. Therefore decreasing the number of function
calls per spike has an increasing benefit.

The picture is less clear for the average number of clockticks required
to complete an instruction (\prettyref{subsec:results_vtune}). For
small networks the new algorithms exhibit an advantage. However, with
increasing network size, eventually more clockticks per instruction
are required than by the original algorithm. Nevertheless, these latencies
are hard to compare as the new algorithm executes only half of the
instructions and may therefore put memory interfaces under larger
stress. This result already indicates that methods of latency hiding
may now be successful in further reducing spike-delivery time. The
product of the number of instructions and the clockticks per instruction
gives an estimate of the total number of clockticks required. The
observed stable improvement across all network sizes confirms the
direct measurements of simulation time.

Faster simulation can trivially be achieved by reducing the generality
of the code or by reducing the accuracy of the simulation. While the
SRR optimization does not touch the code of individual neuron or synapse
models, a critical point in the P2RB optimization with regard to code
generality is the replacement of the target identifier in the synapse
object (\prettyref{subsec:original-deliver-algo}) by a pointer to
the corresponding spike ring buffer. Synaptic plasticity is the biological
phenomenon by which the strength of a synapse changes in dependence
of the spiking activity of the presynaptic and the postsynaptic neuron.
This is one of the key mechanisms by which brains implement system-level
learning. For a wide class of models of synaptic plasticity it is
sufficient to update the synaptic weight when a presynaptic spike
arrives at the synapse \citep{Morrison08_459,stapmanns2020event}.
However, at this point in time the synapse typically needs to inspect
a state variable of the postsynaptic neuron or even retrieve the spiking
history of the postsynaptic neuron since the last presynaptic spike.
This information is only available in the neuron, not in the spike
ring buffer. Still generality is preserved as in the reference simulation
engine \citep{Gewaltig_07_11204} synapses are not restricted to a
single strategy for accessing the target neuron or its spike ring
buffer. A static synapse can implement the P2RB idea while a plastic
synapse stays with the target identifier from which the state of the
neuron as well as the spike ring buffer can be reached. But in this
way a plastic synapse does not profit from the advantages of P2RB
at all. There are two alternatives. First, the spike ring buffer can
be equipped with a pointer to the target neuron. This requires an
indirection in the update of the synapse but still avoids the need
to select the correct ring buffer during spike delivery. Second, the
synapse can store both a target identifier and a pointer to the ring
buffer. This removes the indirection for the price of additional per-synapse
memory usage. There are no fundamental limitations preventing us from
making both solutions available to the neuroscientist via different
synapse types. In fact, this strategy is currently in use, for example,
to provide synapse types with different target identifiers either
consuming less memory or requiring fewer indirections (\prettyref{subsec:fundamental-data-structures}),
where template-based solutions prevent the duplication of entire model
codes. Users can thus select the optimal synapse-type version depending
on the amount of memory available. However, making multiple versions
of the same model available reduces the user-friendliness of the application.
A domain specific language like NESTML \citep{Plotnikov16_93} may
come to rescue here generating more compact or faster code depending
on hints of the neuroscientists to the compiler. This idea could be
extended to other parts of the simulation cycle where further information
is required to decide on a suitable optimization.

The incoming spike events of a compute node specify the hosting thread
as well as the location of the synaptic targets, but they are unsorted
with respect to the hosting thread and synapse type. Nevertheless
the present work shows that the processing of spikes can be completely
parallelized requiring only a single synchronization between the threads
at the point where the spikes are sorted according to target thread
and synapse type, which is when all spikes have been transferred from
the MPI receive buffer into the novel spike-receive register. This
suggests that spike delivery fully profits from a further increase
in number of threads per compute node. Although here we concentrate
on compute nodes with on the order of ten cores per processor we expect
that the benefits of parallelization extend to at least an order of
magnitude more cores, which matches recent hardware developments.
The scaling might still be limited by the structure of the spike-receive
register having separate domains for each thread writing spike data
from the MPI spike receive buffer to the register. If the same number
of spikes is handled by more threads, the spikes are distributed to
more domains of the register such that during the actual delivery
from the register to the thread-local targets each thread needs to
collect its spikes from more memory domains.

The local processing of a compute node is now better understood and
for large networks the communication between nodes begins to dominate
simulation time already for the machines investigated here. Current
chip technology is essentially two-dimensional in contrast to the
three-dimensional organization of the brain and parallelization in
the brain is more fine grained. Inside a compute node technology compensates
for these advantages by communication over buses. After substantially
reducing the number of instructions, we see indications that memory
latency is a problem when spikes from many sources need to be processed.
Therefore it remains to be seen whether techniques of latency hiding
can further push the limits imposed by the von Neumann bottleneck.\textcolor{black}{{}
Any neuromorphic hardware based on compute nodes communicating by
a collective spike exchange in fixed time intervals needs to organize
routing of the spikes to target neurons. The ideas presented in the
present study on streamlining this process by partial parallel sorting
may help in the design of adequate hardware support.} However, between
compute nodes the latency of state-of-the-art inter-node communication
fabrics is likely to be the next limiting factor for simulation. A
possible approach to mitigate this problem is the design of dedicated
neuromorphic hardware explicitly optimized for communication. The
SpiNNaker project \citep{Furber12_1,furberpetrut}, for example, follows
an extreme approach by routing packets with individual spikes to the
respective processing units.

Part of the improvements in performance this study achieves come
at the price of an increase in the number of lines of code and an
increase in code complexity. In general one needs to weigh the achieved
performance improvements against detrimental effects on maintainability.
This is particularly relevant for a community code like the one under
consideration, in which experienced developers are continuously replaced
by new contributors. Highly optimized code may be more difficult to
keep up to date and adjust to future compute node architectures. Next
to conceptual documentation of optimization to core algorithms, code
generation, as explored in the NESTML project \citep{Plotnikov16_93},
may be part of a strategy to reduce this friction between performance
and code accessibility. A domain specific language lets a spectrum
of users concentrate on the formal description of the problem while
experienced developers make sure the generator produces optimized
code, possibly even adapted to specific target architectures. Until
simulations are fast enough to enable the investigation of plastic
networks at natural density we have to find ways to cope with an increasing
complexity of the algorithms and their respective implementations.

Over the last two decades studies on simulation technology for spiking
neuronal networks regularly report improvements in simulation speed
on the order of several percent and improved scaling compared to the
state-of-the-art technology. The stream of publications on simulation
technology in the field shows that there was and still is room for
substantial improvements. Nevertheless, at first sight it seems implausible
that over this time span no canonical algorithm has emerged and progress
shows no sign of saturation. The solution to this riddle is that new
articles tend to immediately concentrate on the latest available hardware
and are interested in their limits in terms of network size. This
is driven by the desire of neuroscience to overcome the limitations
of extremely downscaled models and arrive at a technology capable
of representing relevant parts of the brain. Moreover, investigations
of novel models in computational neuroscience have a life-cycle of
roughly five years, the same time scale at which supercomputers are
installed and decommissioned. Thus, both representative network models
and the hardware to simulate them are in flux, which makes comprehensive
performance studies difficult. The software evolution of spiking
network simulation code is largely unknown and \textcolor{black}{the
community} may profit from a review exposing dead ends and volatile
locations of the algorithm. \textcolor{black}{For a more systematic
monitoring of technological progress} the community needs to learn
how to establish and maintain reference models and \textcolor{black}{keep
track of benchmarking} data \textcolor{black}{and their respective
metadata}.

The present study streamlines the routing of spikes in a compute
node by a fully parallel partial sorting of the incoming spikes and
refactoring of the code. This halves the number of instructions for
this phase of the simulation and leads to a substantial reduction
in simulation time. We expect that our work provides the basis for
the successful application of techniques of latency hiding and vectorization.

\section*{Data Availability Statement}

The datasets presented in this study can be found in \textcolor{black}{the
}\texttt{}online repositor\textcolor{black}{y} \citet{pronold_jari_routing_sorting_code}.

\section*{Code Availability}

The new data structures and algorithms, the code for model simulations,
and Python scripts for the analysis and results are fully available
at: \citet{pronold_jari_routing_sorting_code}.

\section*{Funding}

\phantomsection\addcontentsline{toc}{section}{Funding}

Partly supported by the European Union's Horizon 2020 (H2020) funding
framework under grant agreement no. 785907 (Human Brain Project, HBP
SGA2), no. 945539 (HBP SGA3), and no. 754304 (DEEP-EST), the Helmholtz
Association Initiative and Networking Fund under project number SO-092
(Advanced Computing Architectures, ACA), and the Deutsche Forschungsgemeinschaft
(DFG, German Research Foundation) - 368482240/GRK2416. Use of the
JURECA supercomputer in J\"ulich was made possible through VSR computation
time grant Brain-Scale Simulations JINB33. This research used resources
of K computer at the RIKEN Advanced Institute for Computational Science.
Supported by the project Exploratory Challenge on Post-K Computer
(Understanding the neural mechanisms of thoughts and its applications
to AI) of the Ministry of Education, Culture, Sports, Science and
Technology (MEXT).

\section*{Conflict of Interest}

The authors declare that the research was conducted in the absence
of any commercial or financial relationships that could be construed
as a potential conflict of interest.

\section*{Acknowledgments}

\phantomsection\addcontentsline{toc}{section}{Acknowledgments}

We are grateful to Mitsuhisa Sato for his guidance, which helped
us shape the project, to Johanna Senk and Dennis Terhorst for fruitful
discussions and joint efforts during the HPC Optimisation and Scaling
Workshop 2019 at the Jülich Supercomputing Centre, Germany, to Sebastian
Lührs for his help with JUBE, to our colleagues in the Simulation
and Data Laboratory Neuroscience of the J\"ulich Supercomputing Centre
for continuous collaboration, and to the members of the NEST development
community for their contributions to the concepts and implementation
of the NEST simulator. All network simulations were carried out with
NEST (\href{http://www.nest-simulator.org}{http://www.nest-simulator.org}).

\phantomsection\addcontentsline{toc}{section}{\refname} 

\bibliographystyle{elsarticle-num-names-custom}

\addcontentsline{toc}{section}{\refname}
\begin{thebibliography}{45}
\expandafter\ifx\csname natexlab\endcsname\relax\def\natexlab#1{#1}\fi
\providecommand{\url}[1]{\texttt{#1}}
\providecommand{\href}[2]{#2}
\providecommand{\path}[1]{#1}
\providecommand{\DOIprefix}{doi:}
\providecommand{\ArXivprefix}{arXiv:}
\providecommand{\URLprefix}{URL: }
\providecommand{\Pubmedprefix}{pmid:}
\providecommand{\doi}[1]{\href{http://dx.doi.org/#1}{\path{#1}}}
\providecommand{\Pubmed}[1]{\href{pmid:#1}{\path{#1}}}
\providecommand{\bibinfo}[2]{#2}
\ifx\xfnm\relax \def\xfnm[#1]{\unskip,\space#1}\fi
%Type = Article
\bibitem[{Jordan et~al.(2018)Jordan, Ippen, Helias, Kitayama, Sato, Igarashi,
  Diesmann, and Kunkel}]{Jordan18_2}
\bibinfo{author}{J.~Jordan}, \bibinfo{author}{T.~Ippen},
  \bibinfo{author}{M.~Helias}, \bibinfo{author}{I.~Kitayama},
  \bibinfo{author}{M.~Sato}, \bibinfo{author}{J.~Igarashi},
  \bibinfo{author}{M.~Diesmann}, \bibinfo{author}{S.~Kunkel},
\newblock \bibinfo{title}{Extremely scalable spiking neuronal network
  simulation code: From laptops to exascale computers},
\newblock \bibinfo{journal}{Front. Neuroinformatics} \bibinfo{volume}{12}
  (\bibinfo{year}{2018}) \bibinfo{pages}{2}.
  \DOIprefix\doi{10.3389/fninf.2018.00002}.
%Type = Article
\bibitem[{Schmidt et~al.(2018)Schmidt, Bakker, Hilgetag, Diesmann, and van
  Albada}]{Schmidt18_1409}
\bibinfo{author}{M.~Schmidt}, \bibinfo{author}{R.~Bakker},
  \bibinfo{author}{C.~C. Hilgetag}, \bibinfo{author}{M.~Diesmann},
  \bibinfo{author}{S.~J. van Albada},
\newblock \bibinfo{title}{Multi-scale account of the network structure of
  macaque visual cortex},
\newblock \bibinfo{journal}{Brain Struct. Func.} \bibinfo{volume}{223}
  (\bibinfo{year}{2018}) \bibinfo{pages}{1409--1435}.
  \DOIprefix\doi{10.1007/s00429-017-1554-4}.
%Type = Article
\bibitem[{Morrison et~al.(2005)Morrison, Mehring, Geisel, Aertsen, and
  Diesmann}]{Morrison05a}
\bibinfo{author}{A.~Morrison}, \bibinfo{author}{C.~Mehring},
  \bibinfo{author}{T.~Geisel}, \bibinfo{author}{A.~Aertsen},
  \bibinfo{author}{M.~Diesmann},
\newblock \bibinfo{title}{Advancing the boundaries of high connectivity network
  simulation with distributed computing},
\newblock \bibinfo{journal}{Neural Comput.} \bibinfo{volume}{17}
  (\bibinfo{year}{2005}) \bibinfo{pages}{1776--1801}.
  \DOIprefix\doi{10.1162/0899766054026648}.
%Type = Article
\bibitem[{Markram et~al.(2015)Markram, Muller, Ramaswamy, Reimann, Abdellah,
  Sanchez, Ailamaki, Alonso-Nanclares, Antille, Arsever, Kahou, Berger,
  Bilgili, Buncic, Chalimourda, Chindemi, Courcol, Delalondre, Delattre,
  Druckmann, Dumusc, Dynes, Eilemann, Gal, Gevaert, Ghobril, Gidon, Graham,
  Gupta, Haenel, Hay, Heinis, Hernando, Hines, Kanari, Keller, Kenyon, Khazen,
  Kim, King, Kisvarday, Kumbhar, Lasserre, B{\'{e}}, Magalh{\~{a}}es,
  Merch{\'{a}}n-P{\'{e}}rez, Meystre, Morrice, Muller,
  Mu{\~{n}}oz-C{\'{e}}spedes, Muralidhar, Muthurasa, Nachbaur, Newton, Nolte,
  Ovcharenko, Palacios, Pastor, Perin, Ranjan, Riachi, Rodr{\'{\i}}guez,
  Riquelme, R\"{o}ssert, Sfyrakis, Shi, Shillcock, Silberberg, Silva, Tauheed,
  Telefont, Toledo-Rodriguez, Tr\"{a}nkler, Geit, D{\'{\i}}az, Walker, Wang,
  Zaninetta, DeFelipe, Hill, Segev, and Sch\"{u}rmann}]{Markram2015_456}
\bibinfo{author}{H.~Markram}, \bibinfo{author}{E.~Muller},
  \bibinfo{author}{S.~Ramaswamy}, \bibinfo{author}{M.~W. Reimann},
  \bibinfo{author}{M.~Abdellah}, \bibinfo{author}{C.~A. Sanchez},
  \bibinfo{author}{A.~Ailamaki}, \bibinfo{author}{L.~Alonso-Nanclares},
  \bibinfo{author}{N.~Antille}, \bibinfo{author}{S.~Arsever},
  \bibinfo{author}{G.~A.~A. Kahou}, \bibinfo{author}{T.~K. Berger},
  \bibinfo{author}{A.~Bilgili}, \bibinfo{author}{N.~Buncic},
  \bibinfo{author}{A.~Chalimourda}, \bibinfo{author}{G.~Chindemi},
  \bibinfo{author}{J.-D. Courcol}, \bibinfo{author}{F.~Delalondre},
  \bibinfo{author}{V.~Delattre}, \bibinfo{author}{S.~Druckmann},
  \bibinfo{author}{R.~Dumusc}, \bibinfo{author}{J.~Dynes},
  \bibinfo{author}{S.~Eilemann}, \bibinfo{author}{E.~Gal},
  \bibinfo{author}{M.~E. Gevaert}, \bibinfo{author}{J.-P. Ghobril},
  \bibinfo{author}{A.~Gidon}, \bibinfo{author}{J.~W. Graham},
  \bibinfo{author}{A.~Gupta}, \bibinfo{author}{V.~Haenel},
  \bibinfo{author}{E.~Hay}, \bibinfo{author}{T.~Heinis}, \bibinfo{author}{J.~B.
  Hernando}, \bibinfo{author}{M.~Hines}, \bibinfo{author}{L.~Kanari},
  \bibinfo{author}{D.~Keller}, \bibinfo{author}{J.~Kenyon},
  \bibinfo{author}{G.~Khazen}, \bibinfo{author}{Y.~Kim}, \bibinfo{author}{J.~G.
  King}, \bibinfo{author}{Z.~Kisvarday}, \bibinfo{author}{P.~Kumbhar},
  \bibinfo{author}{S.~Lasserre}, \bibinfo{author}{J.-V.~L. B{\'{e}}},
  \bibinfo{author}{B.~R. Magalh{\~{a}}es},
  \bibinfo{author}{A.~Merch{\'{a}}n-P{\'{e}}rez}, \bibinfo{author}{J.~Meystre},
  \bibinfo{author}{B.~R. Morrice}, \bibinfo{author}{J.~Muller},
  \bibinfo{author}{A.~Mu{\~{n}}oz-C{\'{e}}spedes},
  \bibinfo{author}{S.~Muralidhar}, \bibinfo{author}{K.~Muthurasa},
  \bibinfo{author}{D.~Nachbaur}, \bibinfo{author}{T.~H. Newton},
  \bibinfo{author}{M.~Nolte}, \bibinfo{author}{A.~Ovcharenko},
  \bibinfo{author}{J.~Palacios}, \bibinfo{author}{L.~Pastor},
  \bibinfo{author}{R.~Perin}, \bibinfo{author}{R.~Ranjan},
  \bibinfo{author}{I.~Riachi}, \bibinfo{author}{J.-R. Rodr{\'{\i}}guez},
  \bibinfo{author}{J.~L. Riquelme}, \bibinfo{author}{C.~R\"{o}ssert},
  \bibinfo{author}{K.~Sfyrakis}, \bibinfo{author}{Y.~Shi},
  \bibinfo{author}{J.~C. Shillcock}, \bibinfo{author}{G.~Silberberg},
  \bibinfo{author}{R.~Silva}, \bibinfo{author}{F.~Tauheed},
  \bibinfo{author}{M.~Telefont}, \bibinfo{author}{M.~Toledo-Rodriguez},
  \bibinfo{author}{T.~Tr\"{a}nkler}, \bibinfo{author}{W.~V. Geit},
  \bibinfo{author}{J.~V. D{\'{\i}}az}, \bibinfo{author}{R.~Walker},
  \bibinfo{author}{Y.~Wang}, \bibinfo{author}{S.~M. Zaninetta},
  \bibinfo{author}{J.~DeFelipe}, \bibinfo{author}{S.~L. Hill},
  \bibinfo{author}{I.~Segev}, \bibinfo{author}{F.~Sch\"{u}rmann},
\newblock \bibinfo{title}{Reconstruction and simulation of neocortical
  microcircuitry},
\newblock \bibinfo{journal}{Cell} \bibinfo{volume}{163} (\bibinfo{year}{2015})
  \bibinfo{pages}{456--492}. \DOIprefix\doi{10.1016/j.cell.2015.09.029}.
%Type = Book
\bibitem[{Carnevale and Hines(2006)}]{Carnevale06}
\bibinfo{author}{N.~T. Carnevale}, \bibinfo{author}{M.~L. Hines},
  \bibinfo{title}{The {NEURON} Book}, \bibinfo{publisher}{Cambridge University
  Press}, \bibinfo{address}{Cambridge}, \bibinfo{year}{2006}.
%Type = Article
\bibitem[{Kumbhar et~al.(2019)Kumbhar, Hines, Fouriaux, Ovcharenko, King,
  Delalondre, and Schürmann}]{kumbhar2019coreneuron}
\bibinfo{author}{P.~Kumbhar}, \bibinfo{author}{M.~Hines},
  \bibinfo{author}{J.~Fouriaux}, \bibinfo{author}{A.~Ovcharenko},
  \bibinfo{author}{J.~King}, \bibinfo{author}{F.~Delalondre},
  \bibinfo{author}{F.~Schürmann},
\newblock \bibinfo{title}{Coreneuron : An optimized compute engine for the
  neuron simulator},
\newblock \bibinfo{journal}{Front. Neuroinformatics} \bibinfo{volume}{13}
  (\bibinfo{year}{2019}) \bibinfo{pages}{63}.
  \DOIprefix\doi{10.3389/fninf.2019.00063}.
%Type = Inproceedings
\bibitem[{Akar et~al.(2019)Akar, Cumming, Karakasis, K{\"u}sters, Klijn,
  Peyser, and Yates}]{Akar19_274}
\bibinfo{author}{N.~A. Akar}, \bibinfo{author}{B.~Cumming},
  \bibinfo{author}{V.~Karakasis}, \bibinfo{author}{A.~K{\"u}sters},
  \bibinfo{author}{W.~Klijn}, \bibinfo{author}{A.~Peyser},
  \bibinfo{author}{S.~Yates},
\newblock \bibinfo{title}{Arbor ---a morphologically-detailed neural network
  simulation library for contemporary high-performance computing
  architectures},
\newblock in: \bibinfo{booktitle}{2019 27th Euromicro International Conference
  on Parallel, Distributed and Network-Based Processing (PDP)},
  \bibinfo{organization}{IEEE}, \bibinfo{year}{2019}, pp.
  \bibinfo{pages}{274--282}.
%Type = Article
\bibitem[{Brunel(2000)}]{Brunel00_183}
\bibinfo{author}{N.~Brunel},
\newblock \bibinfo{title}{Dynamics of sparsely connected networks of excitatory
  and inhibitory spiking neurons},
\newblock \bibinfo{journal}{Journal of Computational Neuroscience}
  \bibinfo{volume}{8} (\bibinfo{year}{2000}) \bibinfo{pages}{183--208}.
  \DOIprefix\doi{10.1023/a:1008925309027}.
%Type = Article
\bibitem[{Potjans and Diesmann(2014)}]{Potjans14_785}
\bibinfo{author}{T.~C. Potjans}, \bibinfo{author}{M.~Diesmann},
\newblock \bibinfo{title}{The cell-type specific cortical microcircuit:
  Relating structure and activity in a full-scale spiking network model},
\newblock \bibinfo{journal}{Cereb. Cortex} \bibinfo{volume}{24}
  (\bibinfo{year}{2014}) \bibinfo{pages}{785--806}.
  \DOIprefix\doi{10.1093/cercor/bhs358}.
%Type = Article
\bibitem[{Billeh et~al.(2020)Billeh, Cai, Gratiy, Dai, Iyer, Gouwens,
  Abbasi-Asl, Jia, Siegle, Olsen et~al.}]{Billeh20}
\bibinfo{author}{Y.~N. Billeh}, \bibinfo{author}{B.~Cai},
  \bibinfo{author}{S.~L. Gratiy}, \bibinfo{author}{K.~Dai},
  \bibinfo{author}{R.~Iyer}, \bibinfo{author}{N.~W. Gouwens},
  \bibinfo{author}{R.~Abbasi-Asl}, \bibinfo{author}{X.~Jia},
  \bibinfo{author}{J.~H. Siegle}, \bibinfo{author}{S.~R. Olsen}, et~al.,
\newblock \bibinfo{title}{Systematic integration of structural and functional
  data into multi-scale models of mouse primary visual cortex},
\newblock \bibinfo{journal}{Neuron} \bibinfo{volume}{106}
  (\bibinfo{year}{2020}) \bibinfo{pages}{388--403.e18}.
  \DOIprefix\doi{https://doi.org/10.1016/j.neuron.2020.01.040}.
%Type = Article
\bibitem[{Schmidt et~al.(2018)Schmidt, Bakker, Shen, Bezgin, Diesmann, and van
  Albada}]{Schmidt18_e1006359}
\bibinfo{author}{M.~Schmidt}, \bibinfo{author}{R.~Bakker},
  \bibinfo{author}{K.~Shen}, \bibinfo{author}{G.~Bezgin},
  \bibinfo{author}{M.~Diesmann}, \bibinfo{author}{S.~J. van Albada},
\newblock \bibinfo{title}{A multi-scale layer-resolved spiking network model of
  resting-state dynamics in macaque visual cortical areas},
\newblock \bibinfo{journal}{PLOS Comput. Biol.} \bibinfo{volume}{14}
  (\bibinfo{year}{2018}) \bibinfo{pages}{e1006359}.
  \DOIprefix\doi{10.1371/journal.pcbi.1006359}.
%Type = Article
\bibitem[{Joglekar et~al.(2018)Joglekar, Mejias, Yang, and
  Wang}]{Joglekar18_222}
\bibinfo{author}{M.~R. Joglekar}, \bibinfo{author}{J.~F. Mejias},
  \bibinfo{author}{G.~R. Yang}, \bibinfo{author}{X.-J. Wang},
\newblock \bibinfo{title}{Inter-areal balanced amplification enhances signal
  propagation in a large-scale circuit model of the primate cortex},
\newblock \bibinfo{journal}{Neuron} \bibinfo{volume}{98} (\bibinfo{year}{2018})
  \bibinfo{pages}{222--234}.
%Type = Article
\bibitem[{Cremonesi et~al.(2020)Cremonesi, Hager, Wellein, and
  Sch{\"u}rmann}]{cremonesi2020analytic}
\bibinfo{author}{F.~Cremonesi}, \bibinfo{author}{G.~Hager},
  \bibinfo{author}{G.~Wellein}, \bibinfo{author}{F.~Sch{\"u}rmann},
\newblock \bibinfo{title}{Analytic performance modeling and analysis of
  detailed neuron simulations},
\newblock \bibinfo{journal}{Int. J. High Perform. Comput. Appl.}
  \bibinfo{volume}{34} (\bibinfo{year}{2020}) \bibinfo{pages}{428--449}.
  \DOIprefix\doi{10.1177/1094342020912528}.
%Type = Article
\bibitem[{Cremonesi and Sch{\"u}rmann(2020)}]{cremonesi2020understanding}
\bibinfo{author}{F.~Cremonesi}, \bibinfo{author}{F.~Sch{\"u}rmann},
\newblock \bibinfo{title}{{Understanding Computational Costs of Cellular-Level
  Brain Tissue Simulations Through Analytical Performance Models}},
\newblock \bibinfo{journal}{Neuroinformatics} \bibinfo{volume}{18}
  (\bibinfo{year}{2020}) \bibinfo{pages}{407--428}.
  \DOIprefix\doi{10.1007/s12021-019-09451-w}.
%Type = Article
\bibitem[{Gewaltig and Diesmann(2007)}]{Gewaltig_07_11204}
\bibinfo{author}{M.-O. Gewaltig}, \bibinfo{author}{M.~Diesmann},
\newblock \bibinfo{title}{{NEST} ({NE}ural {S}imulation {T}ool)},
\newblock \bibinfo{journal}{Scholarpedia} \bibinfo{volume}{2}
  (\bibinfo{year}{2007}) \bibinfo{pages}{1430}.
  \DOIprefix\doi{10.4249/scholarpedia.1430}.
%Type = Article
\bibitem[{Kunkel(2019)}]{Kunkel19_ISC}
\bibinfo{author}{S.~Kunkel},
\newblock \bibinfo{title}{Routing brain traffic through the bottlenecks of
  general purpose computers: Challenges for spiking neural network simulation
  code},
\newblock \bibinfo{journal}{ISC} \bibinfo{volume}{33} (\bibinfo{year}{2019}).
  \bibinfo{note}{High Performance Computing: ISC High Performance 2019
  International, Frankfurt, Germany, June 16-20, 2019}.
%Type = Article
\bibitem[{Kunkel et~al.(2014)Kunkel, Schmidt, Eppler, Masumoto, Igarashi,
  Ishii, Fukai, Morrison, Diesmann, and Helias}]{Kunkel14_78}
\bibinfo{author}{S.~Kunkel}, \bibinfo{author}{M.~Schmidt},
  \bibinfo{author}{J.~M. Eppler}, \bibinfo{author}{G.~Masumoto},
  \bibinfo{author}{J.~Igarashi}, \bibinfo{author}{S.~Ishii},
  \bibinfo{author}{T.~Fukai}, \bibinfo{author}{A.~Morrison},
  \bibinfo{author}{M.~Diesmann}, \bibinfo{author}{M.~Helias},
\newblock \bibinfo{title}{Spiking network simulation code for petascale
  computers},
\newblock \bibinfo{journal}{Front. Neuroinformatics} \bibinfo{volume}{8}
  (\bibinfo{year}{2014}) \bibinfo{pages}{78}.
  \DOIprefix\doi{10.3389/fninf.2014.00078}.
%Type = Article
\bibitem[{Morrison et~al.(2007)Morrison, Aertsen, and
  Diesmann}]{Morrison07_1437}
\bibinfo{author}{A.~Morrison}, \bibinfo{author}{A.~Aertsen},
  \bibinfo{author}{M.~Diesmann},
\newblock \bibinfo{title}{Spike-timing dependent plasticity in balanced random
  networks},
\newblock \bibinfo{journal}{Neural Comput.} \bibinfo{volume}{19}
  (\bibinfo{year}{2007}) \bibinfo{pages}{1437--1467}.
  \DOIprefix\doi{10.1162/neco.2007.19.6.1437}.
%Type = Article
\bibitem[{Einevoll et~al.(2019)Einevoll, Destexhe, Diesmann, Gr{\"u}n, Jirsa,
  {de Kamps}, Migliore, Ness, Plesser, and Sch{\"u}rmann}]{Einevoll19_735}
\bibinfo{author}{G.~T. Einevoll}, \bibinfo{author}{A.~Destexhe},
  \bibinfo{author}{M.~Diesmann}, \bibinfo{author}{S.~Gr{\"u}n},
  \bibinfo{author}{V.~Jirsa}, \bibinfo{author}{M.~{de Kamps}},
  \bibinfo{author}{M.~Migliore}, \bibinfo{author}{T.~V. Ness},
  \bibinfo{author}{H.~E. Plesser}, \bibinfo{author}{F.~Sch{\"u}rmann},
\newblock \bibinfo{title}{{The Scientific Case for Brain Simulations}},
\newblock \bibinfo{journal}{Neuron} \bibinfo{volume}{102}
  (\bibinfo{year}{2019}) \bibinfo{pages}{735--744}.
  \DOIprefix\doi{10.1016/j.neuron.2019.03.027}.
%Type = Article
\bibitem[{Morrison et~al.(2007)Morrison, Straube, Plesser, and
  Diesmann}]{Morrison07_47}
\bibinfo{author}{A.~Morrison}, \bibinfo{author}{S.~Straube},
  \bibinfo{author}{H.~E. Plesser}, \bibinfo{author}{M.~Diesmann},
\newblock \bibinfo{title}{Exact subthreshold integration with continuous spike
  times in discrete-time neural network simulations},
\newblock \bibinfo{journal}{Neural Comput.} \bibinfo{volume}{19}
  (\bibinfo{year}{2007}) \bibinfo{pages}{47--79}.
  \DOIprefix\doi{10.1162/neco.2007.19.1.47}.
%Type = Article
\bibitem[{Hanuschkin et~al.(2010)Hanuschkin, Kunkel, Helias, Morrison, and
  Diesmann}]{Hanuschkin10_113}
\bibinfo{author}{A.~Hanuschkin}, \bibinfo{author}{S.~Kunkel},
  \bibinfo{author}{M.~Helias}, \bibinfo{author}{A.~Morrison},
  \bibinfo{author}{M.~Diesmann},
\newblock \bibinfo{title}{A general and efficient method for incorporating
  precise spike times in globally time-driven simulations},
\newblock \bibinfo{journal}{Front. Neuroinformatics} \bibinfo{volume}{4}
  (\bibinfo{year}{2010}) \bibinfo{pages}{113}.
  \DOIprefix\doi{10.3389/fninf.2010.00113}.
%Type = Incollection
\bibitem[{Morrison and Diesmann(2008)}]{Morrison08_267}
\bibinfo{author}{A.~Morrison}, \bibinfo{author}{M.~Diesmann},
\newblock \bibinfo{title}{Maintaining causality in discrete time neuronal
  network simulations},
\newblock in: \bibinfo{editor}{P.~b. Graben}, \bibinfo{editor}{C.~Zhou},
  \bibinfo{editor}{M.~Thiel}, \bibinfo{editor}{J.~Kurths} (Eds.),
  \bibinfo{booktitle}{Lectures in Supercomputational Neurosciences: Dynamics in
  Complex Brain Networks}, \bibinfo{publisher}{Springer},
  \bibinfo{address}{Berlin, Heidelberg}, \bibinfo{year}{2008}, pp.
  \bibinfo{pages}{267--278}. \DOIprefix\doi{10.1007/978-3-540-73159-7_10}.
%Type = Article
\bibitem[{Helias et~al.(2012)Helias, Kunkel, Masumoto, Igarashi, Eppler, Ishii,
  Fukai, Morrison, and Diesmann}]{Helias12_26}
\bibinfo{author}{M.~Helias}, \bibinfo{author}{S.~Kunkel},
  \bibinfo{author}{G.~Masumoto}, \bibinfo{author}{J.~Igarashi},
  \bibinfo{author}{J.~M. Eppler}, \bibinfo{author}{S.~Ishii},
  \bibinfo{author}{T.~Fukai}, \bibinfo{author}{A.~Morrison},
  \bibinfo{author}{M.~Diesmann},
\newblock \bibinfo{title}{Supercomputers ready for use as discovery machines
  for neuroscience},
\newblock \bibinfo{journal}{Front. Neuroinformatics} \bibinfo{volume}{6}
  (\bibinfo{year}{2012}) \bibinfo{pages}{26}.
  \DOIprefix\doi{10.3389/fninf.2012.00026}.
%Type = Article
\bibitem[{Kunkel et~al.(2012)Kunkel, Potjans, Eppler, Plesser, Morrison, and
  Diesmann}]{Kunkel2012_5_35}
\bibinfo{author}{S.~Kunkel}, \bibinfo{author}{T.~C. Potjans},
  \bibinfo{author}{J.~M. Eppler}, \bibinfo{author}{H.~E. Plesser},
  \bibinfo{author}{A.~Morrison}, \bibinfo{author}{M.~Diesmann},
\newblock \bibinfo{title}{Meeting the memory challenges of brain-scale
  simulation},
\newblock \bibinfo{journal}{Front. Neuroinformatics} \bibinfo{volume}{5}
  (\bibinfo{year}{2012}) \bibinfo{pages}{35}.
  \DOIprefix\doi{10.3389/fninf.2011.00035}.
%Type = Article
\bibitem[{Ippen et~al.(2017)Ippen, Eppler, Plesser, and
  Diesmann}]{Ippen2017_30}
\bibinfo{author}{T.~Ippen}, \bibinfo{author}{J.~M. Eppler},
  \bibinfo{author}{H.~E. Plesser}, \bibinfo{author}{M.~Diesmann},
\newblock \bibinfo{title}{Constructing neuronal network models in massively
  parallel environments},
\newblock \bibinfo{journal}{Front. Neuroinformatics} \bibinfo{volume}{11}
  (\bibinfo{year}{2017}) \bibinfo{pages}{30}.
  \DOIprefix\doi{10.3389/fninf.2017.00030}.
%Type = Article
\bibitem[{Eppler et~al.(2009)Eppler, Helias, Muller, Diesmann, and
  Gewaltig}]{Eppler09_12}
\bibinfo{author}{J.~M. Eppler}, \bibinfo{author}{M.~Helias},
  \bibinfo{author}{E.~Muller}, \bibinfo{author}{M.~Diesmann},
  \bibinfo{author}{M.~Gewaltig},
\newblock \bibinfo{title}{{PyNEST}: a convenient interface to the {NEST}
  simulator},
\newblock \bibinfo{journal}{Front. Neuroinformatics} \bibinfo{volume}{2}
  (\bibinfo{year}{2009}) \bibinfo{pages}{12}.
  \DOIprefix\doi{10.3389/neuro.11.012.2008}.
%Type = Article
\bibitem[{Zaytsev and Morrison(2014)}]{Zaytsev14_23}
\bibinfo{author}{Y.~V. Zaytsev}, \bibinfo{author}{A.~Morrison},
\newblock \bibinfo{title}{{CyNEST: a maintainable Cython-based interface for
  the NEST simulator}},
\newblock \bibinfo{journal}{Front. Neuroinformatics} \bibinfo{volume}{8}
  (\bibinfo{year}{2014}). \DOIprefix\doi{10.3389/fninf.2014.00023}.
%Type = Inproceedings
\bibitem[{Plotnikov et~al.(2016)Plotnikov, Blundell, Ippen, Eppler, Rumpe, and
  Morrison}]{Plotnikov16_93}
\bibinfo{author}{D.~Plotnikov}, \bibinfo{author}{I.~Blundell},
  \bibinfo{author}{T.~Ippen}, \bibinfo{author}{J.~M. Eppler},
  \bibinfo{author}{B.~Rumpe}, \bibinfo{author}{A.~Morrison},
\newblock \bibinfo{title}{{NESTML}: a modeling language for spiking neurons},
\newblock in: \bibinfo{editor}{A.~Oberweis}, \bibinfo{editor}{R.~Reussner}
  (Eds.), \bibinfo{booktitle}{Modellierung 2016}, volume
  \bibinfo{volume}{P-254} of \textit{\bibinfo{series}{Lecture Notes in
  Informatics (LNI)}}, \bibinfo{publisher}{Gesellschaft f\"ur Informatik e.V.
  (GI)}, \bibinfo{year}{2016}, pp. \bibinfo{pages}{93--108}. \URLprefix
  \url{http://juser.fz-juelich.de/record/826510}.
%Type = Article
\bibitem[{Hahne et~al.(2015)Hahne, Helias, Kunkel, Igarashi, Bolten, Frommer,
  and Diesmann}]{Hahne15_00022}
\bibinfo{author}{J.~Hahne}, \bibinfo{author}{M.~Helias},
  \bibinfo{author}{S.~Kunkel}, \bibinfo{author}{J.~Igarashi},
  \bibinfo{author}{M.~Bolten}, \bibinfo{author}{A.~Frommer},
  \bibinfo{author}{M.~Diesmann},
\newblock \bibinfo{title}{A unified framework for spiking and gap-junction
  interactions in distributed neuronal network simulations.},
\newblock \bibinfo{journal}{Front. Neuroinformatics} \bibinfo{volume}{9}
  (\bibinfo{year}{2015}). \DOIprefix\doi{10.3389/fninf.2015.00022}.
%Type = Article
\bibitem[{Jordan et~al.(2020)Jordan, , Helias, Diesmann, and
  Kunkel}]{Jordan20_12}
\bibinfo{author}{J.~Jordan}, , \bibinfo{author}{M.~Helias},
  \bibinfo{author}{M.~Diesmann}, \bibinfo{author}{S.~Kunkel},
\newblock \bibinfo{title}{Efficient communication in distributed simulations of
  spiking neuronal networks with gap junctions},
\newblock \bibinfo{journal}{Front. Neuroinformatics} \bibinfo{volume}{14}
  (\bibinfo{year}{2020}) \bibinfo{pages}{12}.
  \DOIprefix\doi{10.3389/fninf.2020.00012}.
%Type = Article
\bibitem[{Potjans et~al.(2010)Potjans, Morrison, and
  Diesmann}]{Potjans10_103389}
\bibinfo{author}{W.~Potjans}, \bibinfo{author}{A.~Morrison},
  \bibinfo{author}{M.~Diesmann},
\newblock \bibinfo{title}{Enabling functional neural circuit simulations with
  distributed computing of neuromodulated plasticity},
\newblock \bibinfo{journal}{Front. Comput. Neurosci.} \bibinfo{volume}{4}
  (\bibinfo{year}{2010}). \DOIprefix\doi{10.3389/fncom.2010.00141}.
%Type = Article
\bibitem[{Stapmanns et~al.(2021)Stapmanns, Hahne, Helias, Bolten, Diesmann, and
  Dahmen}]{Stapmanns21_609147}
\bibinfo{author}{J.~Stapmanns}, \bibinfo{author}{J.~Hahne},
  \bibinfo{author}{M.~Helias}, \bibinfo{author}{M.~Bolten},
  \bibinfo{author}{M.~Diesmann}, \bibinfo{author}{D.~Dahmen},
\newblock \bibinfo{title}{Event-based update of synapses in voltage-based
  learning rules},
\newblock \bibinfo{journal}{Front. Neuroinformatics} \bibinfo{volume}{15}
  (\bibinfo{year}{2021}) \bibinfo{pages}{609147}.
%Type = Article
\bibitem[{Diaz-Pier et~al.(2016)Diaz-Pier, Naveau, Butz-Ostendorf, and
  Morrison}]{Diaz16_57}
\bibinfo{author}{S.~Diaz-Pier}, \bibinfo{author}{M.~Naveau},
  \bibinfo{author}{M.~Butz-Ostendorf}, \bibinfo{author}{A.~Morrison},
\newblock \bibinfo{title}{Automatic generation of connectivity for large-scale
  neuronal network models through structural plasticity},
\newblock \bibinfo{journal}{Front. Neuroanatomy} \bibinfo{volume}{10}
  (\bibinfo{year}{2016}) \bibinfo{pages}{57}.
  \DOIprefix\doi{10.3389/fnana.2016.00057}.
%Type = Article
\bibitem[{Grytskyy et~al.(2013)Grytskyy, Tetzlaff, Diesmann, and
  Helias}]{Grytskyy13_131}
\bibinfo{author}{D.~Grytskyy}, \bibinfo{author}{T.~Tetzlaff},
  \bibinfo{author}{M.~Diesmann}, \bibinfo{author}{M.~Helias},
\newblock \bibinfo{title}{A unified view on weakly correlated recurrent
  networks},
\newblock \bibinfo{journal}{Front. Comput. Neurosci.} \bibinfo{volume}{7}
  (\bibinfo{year}{2013}) \bibinfo{pages}{131}.
  \DOIprefix\doi{10.3389/fncom.2013.00131}.
%Type = Article
\bibitem[{Hahne et~al.(2017)Hahne, Dahmen, Schuecker, Frommer, Bolten, Helias,
  and Diesmann}]{Hahne17_34}
\bibinfo{author}{J.~Hahne}, \bibinfo{author}{D.~Dahmen},
  \bibinfo{author}{J.~Schuecker}, \bibinfo{author}{A.~Frommer},
  \bibinfo{author}{M.~Bolten}, \bibinfo{author}{M.~Helias},
  \bibinfo{author}{M.~Diesmann},
\newblock \bibinfo{title}{Integration of continuous-time dynamics in a spiking
  neural network simulator},
\newblock \bibinfo{journal}{Front. Neuroinformatics} \bibinfo{volume}{11}
  (\bibinfo{year}{2017}) \bibinfo{pages}{34}.
  \DOIprefix\doi{10.3389/fninf.2017.00034}.
%Type = Article
\bibitem[{Krause and Th{\"o}rnig(2018)}]{Krause:850758}
\bibinfo{author}{D.~Krause}, \bibinfo{author}{P.~Th{\"o}rnig},
\newblock \bibinfo{title}{{JURECA}: {M}odular supercomputer at {J}{\"u}lich
  {S}upercomputing {C}entre},
\newblock \bibinfo{journal}{Journal of large-scale research facilities}
  \bibinfo{volume}{4} (\bibinfo{year}{2018}) \bibinfo{pages}{A132}.
  \DOIprefix\doi{10.17815/jlsrf-4-121-1}.
%Type = Article
\bibitem[{Miyazaki et~al.(2012)Miyazaki, Kusano, Shinjou, Fumiyoshi, Yokokawa,
  and Watanabe}]{Miyazaki12}
\bibinfo{author}{H.~Miyazaki}, \bibinfo{author}{Y.~Kusano},
  \bibinfo{author}{N.~Shinjou}, \bibinfo{author}{S.~Fumiyoshi},
  \bibinfo{author}{M.~Yokokawa}, \bibinfo{author}{T.~Watanabe},
\newblock \bibinfo{title}{{Overview of the K computer System}},
\newblock \bibinfo{journal}{Fujitsu Scientific and Technical Journal}
  \bibinfo{volume}{48} (\bibinfo{year}{2012}) \bibinfo{pages}{255--265}.
%Type = Inproceedings
\bibitem[{Schenck et~al.(2014)Schenck, Adinetz, Zaytsev, Pleiter, and
  Morrison}]{Schenck14_SC}
\bibinfo{author}{W.~Schenck}, \bibinfo{author}{A.~V. Adinetz},
  \bibinfo{author}{Y.~V. Zaytsev}, \bibinfo{author}{D.~Pleiter},
  \bibinfo{author}{A.~Morrison},
\newblock \bibinfo{title}{Performance model for large--scale neural simulations
  with {N}{E}{S}{T}},
\newblock in: \bibinfo{booktitle}{Extended Poster Abstracts of the SC14
  Conference for Supercomputing}, \bibinfo{address}{New Orleans (LA)},
  \bibinfo{year}{2014}.
%Type = Techreport
\bibitem[{Cremonesi(2019)}]{cremonesi2019computational}
\bibinfo{author}{F.~Cremonesi}, \bibinfo{title}{Computational characteristics
  and hardware implications of brain tissue simulations},
  \bibinfo{type}{Technical Report}, EPFL, \bibinfo{year}{2019}.
  \DOIprefix\doi{10.5075/epfl-thesis-9767}.
%Type = Inproceedings
\bibitem[{L\"{u}hrs et~al.(2016)L\"{u}hrs, Rohe, Schnurpfeil, Thust, and
  Frings}]{Luehrs16_432}
\bibinfo{author}{S.~L\"{u}hrs}, \bibinfo{author}{D.~Rohe},
  \bibinfo{author}{A.~Schnurpfeil}, \bibinfo{author}{K.~Thust},
  \bibinfo{author}{W.~Frings},
\newblock \bibinfo{title}{{F}lexible and {G}eneric {W}orkflow {M}anagement},
\newblock in: \bibinfo{booktitle}{Parallel Computing: On the Road to Exascale},
  volume~\bibinfo{volume}{27} of \textit{\bibinfo{series}{Advances in parallel
  computing}}, \bibinfo{publisher}{IOS Press}, \bibinfo{address}{Amsterdam},
  \bibinfo{year}{2016}, pp. \bibinfo{pages}{431--438}.
  \DOIprefix\doi{10.3233/978-1-61499-621-7-431}.
%Type = Article
\bibitem[{Morrison et~al.(2008)Morrison, Diesmann, and
  Gerstner}]{Morrison08_459}
\bibinfo{author}{A.~Morrison}, \bibinfo{author}{M.~Diesmann},
  \bibinfo{author}{W.~Gerstner},
\newblock \bibinfo{title}{Phenomenological models of synaptic plasticity based
  on spike-timing},
\newblock \bibinfo{journal}{Biol. Cybern.} \bibinfo{volume}{98}
  (\bibinfo{year}{2008}) \bibinfo{pages}{459--478}.
  \DOIprefix\doi{10.1007/s00422-008-0233-1}.
%Type = Article
\bibitem[{Stapmanns et~al.(2020)Stapmanns, Hahne, Helias, Bolten, Diesmann, and
  Dahmen}]{stapmanns2020event}
\bibinfo{author}{J.~Stapmanns}, \bibinfo{author}{J.~Hahne},
  \bibinfo{author}{M.~Helias}, \bibinfo{author}{M.~Bolten},
  \bibinfo{author}{M.~Diesmann}, \bibinfo{author}{D.~Dahmen},
\newblock \bibinfo{title}{Event-based update of synapses in voltage-based
  learning rules}  (\bibinfo{year}{2020}).
  \href{http://arxiv.org/abs/2009.08667}{{\tt arXiv:2009.08667}}.
%Type = Article
\bibitem[{Furber et~al.(2013)Furber, Lester, Plana, Garside, Painkras, Temple,
  and Brown}]{Furber12_1}
\bibinfo{author}{S.~Furber}, \bibinfo{author}{D.~Lester},
  \bibinfo{author}{L.~Plana}, \bibinfo{author}{J.~Garside},
  \bibinfo{author}{E.~Painkras}, \bibinfo{author}{S.~Temple},
  \bibinfo{author}{A.~Brown},
\newblock \bibinfo{title}{{Overview of the SpiNNaker System Architecture}},
\newblock \bibinfo{journal}{IEEE Trans. Comp.} \bibinfo{volume}{62}
  (\bibinfo{year}{2013}) \bibinfo{pages}{2454--2467}.
  \DOIprefix\doi{10.1109/TC.2012.142}.
%Type = Book
\bibitem[{Furber and Bogdan(2020)}]{furberpetrut}
\bibinfo{author}{S.~Furber}, \bibinfo{author}{P.~Bogdan},
  \bibinfo{title}{SpiNNaker: A Spiking Neural Network Architecture},
  \bibinfo{publisher}{Boston-Delft: now publishers}, \bibinfo{year}{2020}.
  \DOIprefix\doi{10.1561/9781680836523}.
%Type = Misc
\bibitem[{Pronold et~al.(2021)Pronold, Jordan, Wylie, Kitayama, Diesmann, and
  Kunkel}]{pronold_jari_routing_sorting_code}
\bibinfo{author}{J.~Pronold}, \bibinfo{author}{J.~Jordan},
  \bibinfo{author}{B.~Wylie}, \bibinfo{author}{I.~Kitayama},
  \bibinfo{author}{M.~Diesmann}, \bibinfo{author}{S.~Kunkel},
  \bibinfo{title}{{Code for "Routing brain traffic through the von Neumann
  bottleneck: Parallel sorting and refactoring"}}, \bibinfo{year}{2021}.
  \DOIprefix\doi{10.5281/zenodo.5148731}.

\end{thebibliography}

\end{document}